%% file: amrnewest.tex
\documentclass[12pt]{article}
\usepackage{amsmath}
\usepackage{amsfonts}
\usepackage{epsfig}
\usepackage{color}
\usepackage{float}
\usepackage{amssymb}
\usepackage{comment}
\usepackage{a4wide}
\usepackage{times}[20]
\usepackage{mathptmx}
\newtheorem{thm}{Theorem}[section]

\newtheorem{lemma}{Lemma}[section]

\def\FF{\hbox to 8.33887pt{\rm I\hskip-1.8pt F}}
\def\NN{\hbox to 9.3111pt{\rm I\hskip-1.8pt N}}
\def\PP{\hbox to 8.61664pt{\rm I\hskip-1.8pt P}}
\def\QQ{\rlap {\raise 0.4ex \hbox{$\scriptscriptstyle |$}}
{\hskip -4.5pt Q}}
\def\RR{\hbox to 9.1722pt{\rm I\hskip-1.8pt R}}
\def\ZZ{\hbox to 8.2222pt{\rm Z\hskip-4pt \rm Z}}

\newcommand{\resetequ}{\setcounter{equation}{0}}

\newcommand{\be}{\begin{equation}}
\newcommand{\ee}{\end{equation}}
\newcommand{\bqa}{\begin{eqnarray}}
\newcommand{\eqa}{\end{eqnarray}}
\newcommand{\ba}{\begin{array}}
\newcommand{\ea}{\end{array}}

\newcommand{\qed}{\hfill \rule {1ex}{1ex}}
\newcommand{\al}{\alpha}

\newcommand{\ch}{\chi}

\newcommand{\la}{\lambda}

\newcommand{\si}{\sigma}

\newcommand{\Si}{\Sigma}

\title{Renormalization of the 2-point function\\ of the Hubbard model at
  half-filling}
\author{St\'ephane Afchain, Jacques Magnen, \\  Centre de Physique Th\'eorique, CNRS, UMR 7644, \'Ecole Polytechnique, \\ 91128 Palaiseau - France \\ Vincent Rivasseau \\ Laboratoire de Physique Th\'eorique, CNRS, UMR 8627\\
Universit\'e de Paris-Sud, 91405 Orsay}

\begin{document}

\maketitle

\begin{abstract} We prove that the Hubbard model at finite temperature $T$ and
half-filling is analytic in its coupling constant $\lambda$ 
for $|\lambda| \le  c/|\log T|^{2}$, where $c$ is some numerical constant. 
We also bound the self-energy and prove that the Hubbard model at half-filling is {\it not} a 
Fermi liquid (in the mathematically precise sense of Salmhofer), modulo 
a simple lower bound on the first non-trivial self-energy graph, which will be published
in a companion paper.
\end{abstract}

\resetequ
\section{Introduction}

In \cite{R1} we introduced the tools for a multiscale analysis 
of the two dimensional Hubbard model at half-filling: momentum 
slices, sectors and their conservation rules.

In this paper we achieve the proof that the correlation functions
of the model at finite temperature $T$ are analytic in the coupling constant $\lambda$ 
for $|\lambda| \le  c/|\log T|^{2}$, by treating the renormalization of 
``bipeds'' (two-particle subgraphs), that was missing in \cite{R1}.

This proof requires a new tool which is a constructive two-particle irreducible
analysis of the self-energy. This analysis according to the {\it line form}
of Menger's theorem (\cite{Bondy})
leads to the explicit construction of three line-disjoint paths for every self-energy contribution,
in a way compatible with constructive bounds. On top of that analysis, another one which is 
scale-dependent is performed:
after reduction of some maximal
subsets provided by the scale analysis, two vertex-disjoint
paths are selected in every self-energy contribution. This requires 
a second use of Menger's theorem, now in the {\it vertex form}.
This construction allows to improve the power counting
for two point subgraphs, exploiting the particle-hole symmetry of the theory at half-filling,
and leads to our analyticity result.

In the last section we write the upper bounds on the self-energy
that follow from our analysis. These upper bounds strongly suggest
that the second momentum derivative of the self energy
is not uniformly bounded in the region $|\lambda| \le  c/|\log T|^{2}$.
A rigorous proof of this last statement follows form a rigorous {\it lower} 
bound of the same type than these upper bounds, but for the smallest
non trivial self-energy graph, so as to rule out any "miraculous cancellation". 
This lower bound, which we have now completed, is the tedious but rather 
straightforward study of a single finite dimensional integral. 
Since it is not related to the main analysis 
in this paper, we postpone it to a separate publication \cite{AMR1}.

Taken all together these bounds prove that 
the model is {\it not} a Fermi liquid in the sense of Salmhofer's criterion (see \cite{S1} and \cite{Salm}). Indeed to be such a Fermi liquid the second derivative would have to be 
uniformly bounded in a {\it larger} region  (of type $|\lambda| \le  c/|\log T|$)
than the one for which we prove it is unbounded. 
The scaling properties of the self energy and its derivatives in fact
mean that the model is not of Fermi but of Luttinger type, with logarithmic corrections
if we compare to the standard one dimensional Luttinger liquid.

Let us state precisely the main result of this paper:

\medskip
{\bf Theorem :}
{\it
The radius of convergence of the Hubbard model perturbative series at half-filling is
at least $c/\log^2 T$, where $T$ is the temperature and $c$ some numerical constant.
As $T$ and $\lambda$ jointly tend to 0 in this domain, the self-energy of the model 
does not display the properties of a  Fermi liquid in the sense of \cite{S1},
but those of a Luttinger liquid (with logarithmic corrections).}  
\medskip

Let us also put our paper in perspective and relation
with other programs of rigorous mathematical study
of interacting Fermi systems. Recall that in dimension 1 there is neither
superconductivity nor extended Fermi surface, and Fermion systems
have been proved to exhibit Luttinger liquid behavior \cite{BG}.
The initial goal of the studies in two or three dimensions was
to understand the low temperature phase of these systems, and in particular to build a rigorous constructive BCS theory of superconductivity.
The mechanism for the formation of Cooper pairs 
and the main technical tool to use (namely the corresponding $1/N$ expansion, where $N$
is the number of sectors which proliferate near the Fermi surface at low temperatures) 
have been identified \cite{FMRT1}. 
But the goal of building a completely rigorous BCS theory {\it ab initio} 
remains elusive because of the technicalities
involved with the constructive control of continuous symmetry breaking. 

So the initial goal was replaced with a more modest one, still
important in view of the controversies over the nature of
two dimensional "Fermi liquids" \cite{And}, namely the rigorous control of what occurs
before pair formation. The last decade has seen excellent progress in this direction. 

As is well known, sufficiently high magnetic field or temperature
are the two different ways to break the Cooper pairs and prevent superconductivity.
Accordingly two approaches were devised for the construction of "Fermi liquids". One is based 
on the use of non-parity invariant Fermi surfaces to prevent pair formation. These surfaces occur physically when generic magnetic fields are applied to two dimensional Fermi systems.  The other is based on Salmhofer's criterion \cite{S1}, in which temperature is the cutoff which prevents pair formation.

In a large series of papers \cite{FKT},
the construction of two dimensional Fermi liquids 
for a wide class of non-parity invariant Fermi surfaces has been completed in great detail by  Feldman, Kn\"orrer and Trubowitz. These papers establish
Fermi liquid behavior in the traditional sense of physics textbooks, namely as a jump
of the density of states at the Fermi surface at zero temperature, but they do not 
apply to the simplest Fermi surfaces, such as circles or squares, which are parity invariant.

An other program in recent years was to explore which models satisfy 
Salmhofer's criterion. Of particular interest to us are the three most "canonical" models 
in more than one dimension namely:

\begin{itemize}
\item the jellium model in two dimensions, with circular Fermi surface, nicknamed $J_2$, 

\item the half-filled Hubbard model in two dimensions, with square Fermi surface, nicknamed $H_2$,

\item  and the jellium model in three dimensions, with spherical Fermi surface, nicknamed $J_3$.

\end{itemize}

The study of each model has been divided into two main steps of roughly equal difficulty, the control of convergent contributions and the renormalization of the two point functions. 
In this sense, five of the six steps of our program are now completed. $J_2$ is a Fermi liquid  in the sense of Salmhofer \cite{DR1} - \cite{DR2},  $H_2$ is not, and is a Luttinger liquid with logarithmic corrections, according to \cite{R1}, to the present paper, and to \cite{AMR1}. 

Results similar to \cite{DR1} - \cite{DR2}
have been also obtained for more general convex curves not necessarily rotation invariant
such as those of the Hubbard model at low filling, where the Fermi surface
becomes more and more circular, including an improved treatment of the four point functions leading to 
better constants \cite{BGM}.
Therefore as the filling factor of the Hubbard
model is moved from half-filling to low filling, we conclude that
there must be a crossover from Luttinger liquid
behavior to Fermi liquid behavior. This solves the controversy \cite{And} over the
Luttinger or Fermi nature of two-dimensional many-Fermion systems above their critical 
temperature. The short answer is that it depends on the shape of the Fermi surface.

Up to now only the convergent contributions of $J_3$, which is almost certainly a Fermi liquid, have been controlled \cite{DMR}. The renormalization of the two point functions for $J_3$, the last sixth of our program, remains still to be done. This last part is difficult since the 
cutoffs required in \cite{DMR} do not conserve momentum. This means that the two point 
functions that have to be renormalized in this formalism are not automatically one particle irreducible,
as is the case both in \cite{DR2} and in this paper. This complicates their analysis.

\section{Slices, sectors, propagator decay and momentum conservation}

We recall here some generalities that were explained in \cite{R1}, in order to make this paper self-contained. Given a temperature $T > 0$, the Hubbard model lives on $\left[ - \beta, \ \beta \right[ \times \mathbb{Z}^2$, where $\beta = \frac{1}{T}$. Indeed, the real interval $\left[ - \beta, \ \beta \right[$ should be thought of as the circle of radius $\beta$. A generic element of $\left[ - \beta, \ \beta \right[ \times \mathbb{Z}^2$ will be denoted $x = (x_0, \overrightarrow{x})$, where $x_0 \in \left[ - \beta, \ \beta \right[$ and $\overrightarrow{x} = (n_1, n_2) \ \in \mathbb{Z}^2$.

Like in every Fermionic model, the propagator $C(x_0, \overrightarrow{x})$ \footnote{Indeed, the propagator should be seen as depending on two variables $x, \ y \in \left[ - \beta , \ \beta \right[ \times \mathbb{Z}^2$, but by translational invariance, we have $C(x, \ y) = C(0, \ y - x)$ and we shall write in the following simply $C(x)$ instead of $C(0, \ x)$.} is antiperiodic in the variable $x_0$, with antiperiod $\frac{1}{T}$. Therefore, for the Fourier transform of the propagator $\hat{C} (k_0, \overrightarrow{k})$, the relevant values for $k_0$ are discrete and called the Matsubara frequencies~:
\begin{equation}
k_0 = \left( \frac{\pi}{\beta} \right) (2n + 1), \ n \in \mathbb{Z},
\label{Matsubara}
\end{equation}
whereas the vector $\overrightarrow{k}$ lives on the two-dimensional torus $\mathbb{R}^2 / (2 \pi \mathbb{Z})^2$. 

At half-filling and finite temperature $T$, we have~:
\begin{equation}
\hat{C}_{a, b} (k) = \delta_{a, b} \frac{1}{ik_0 - e(\overrightarrow{k})} \ , 
\end{equation}
with $e(\overrightarrow{k}) = \cos \ k_1 + \cos \ k_2$. $a$ and $b$ are spin indices (elements of the set $\{ \uparrow, \ \downarrow \}$), and may sometimes be dropped when they are not essential. Hence the expression of the real space propagator is~:
\be
C_{a, b}(x) =\frac{1}{(2\pi)^2\beta}\; \sum_{k_0} \; 
\int_{-\pi}^{\pi} dk_1 \int_{-\pi}^{\pi} dk_2 \  e^{ik.x}\;
\hat{C}_{a, b}(k) \ .
\label{tfprop}\ee

The notation $\sum_{k_0}$ really means the discrete sum over the integer
$n$ in (\ref{Matsubara}).
When $T \to 0^+$ (which means $\beta\to + \infty$), $k_0$
becomes a continuous variable, the corresponding discrete sum becomes an
integral, and the corresponding propagator $C_{0}(x)$ becomes singular
on the Fermi surface
defined by $k_0=0$ and $e(\vec{k})=0$. 
This Fermi surface is a square of side size $\sqrt{2} \pi$  
(in the first Brillouin zone) joining the corners $(\pm \pi , 0), (0,\pm\pi )$.
We call this square the Fermi square, its faces  and corners are called
the Fermi faces and corners. Considering the periodic boundary conditions,
there are really four Fermi faces, but only two Fermi corners.

In the following, to simplify notations, we will write:
\be
\int d^3k \; \equiv \; \frac{1}{\beta} \sum_{k_0} \int_{[-\pi,\pi]^2} dk_1 \ dk_2
\quad , \quad 
\int d^3x \; \equiv \; {1\over 2}
\int_{-\beta}^{\beta}dx_0 \sum_{\vec{x}\in \mathbb{Z}^2} \ . \label{convention}
\ee

The interaction of the Hubbard model is simply
\be
S_V = \la  \int_V d^3x \  \left( \sum_{a \in \{ \uparrow, \downarrow \}} \overline{\psi}_a (x) \psi_a (x) \right)^2  \label{int} \ ,
\ee
where  $V:= [-\beta,\beta[\times V'$ and $ V'$ is an auxiliary volume
cutoff
in two dimensional space, that will be sent to infinity eventually.
Remark that in (\ref{Matsubara}) $|k_0|\geq \pi/\beta\neq 0$ 
hence the denominator in $C(k)$ can never be 0 at non zero temperature.
This is why the temperature provides a natural infrared cut-off.

We use in this paper  the same slices and sectors than in \cite{R1} and recall the main points for completeness. 
Introducing a fixed number $M>1$, we perform a slice analysis according to
geometric scales of ratio $M$. Like in \cite{R1} since we have a
finite temperature, this analysis should stop for a scale
$ i_{max}(T)$ such that $M^{-i_{max}(T)}\simeq 1/T$. We write simply $i_{max}$
for $i_{max}(T)$.

As in \cite{R1} we use the tilted orthogonal 
basis in momentum space $(e_+, e_-)$,
defined by $e_+ = (1/2)(\pi, \pi)$ and $e_- = (1/2)(-\pi, \pi)$.
In the corresponding coordinates $(k_+, k_{-})$
the Fermi surface is given by $k_+ = \pm 1$ or  $k_- = \pm 1$.
This follows from the identity
\be \cos k_1 + \cos k_2  = 2 \cos \left(\frac{\pi  k_{+}}{2} \right)
\cos \left(\frac{\pi  k_{-}}{2} \right) \ .
\ee
We also use the convenient notations 
 \begin{equation}
 q_{\pm} = k_{\pm} -1 \ {\rm if} \ k_{\pm} \ge 0\ ; \
 q_{\pm} = k_{\pm} +1 \ {\rm if} \ k_{\pm} < 0  \label{qplusorminus}
\end{equation}
so that $0\le | q_{\pm} |   \le 1$.

Picking a Gevrey compact support function $u(r)\in{\cal C}_{0}^\infty(\mathbb{R})$ 
of order $\alpha<1$ which satisfies:
\be
u(r)= 0 \quad {\rm for} \ |r|> 2 \ ; \ u(r) =1
\quad {\rm for} \   |r|<1  \ \ , \label{gevrey}
\ee
we consider the partition of unity:
\be
1 = \sum_{i= 0}^{i_{max}(T) } u_{i}\left( (k_0)^2  +  4 \cos^2 \left(\frac{\pi  k_{+}}{2} \right) \cos^2 \left(\frac{\pi  k_{-}}{2} \right) \right)\ , 
\ee
with
\be
\begin{cases} 
u_{0}(r) =1- u(r) \ ,\\
u_{i}(r)= u \big(M^{2(i-1)}r \big) - u \big(M^{2i}r \big) \ \text{for} \ i \geq 1. \\
\end{cases}
\ee
The sum over $i$ a priori runs from 0 to $+\infty$ to create a partition of unity, but in fact 
since $k_0^2$ is at least of order $M^{-2 i_{max}(T)}$, the sum over $i$ stops as
$i_{max}(T)$. 
This is similar to \cite{R1}. 

The $i$ slice propagator $C_i(k) = C(k) u_i (k)$   is further sliced into the 
$\pm$ directions exactly as in \cite{R1}:

\be  C_i (k) = \sum_{\si  = (i, s_+ , s_-)}C_{\si} (k) \ ,
\ee
where 
\be
C_{\si} (k)  = C_i (k)  v_{s_{+}} \left[\cos ^2 \left(\frac{\pi k_+ }{2} \right)\right]v_{s_{-}} \left[\cos ^2 \left(\frac{\pi k_- }{2} \right)\right]
\ee
using a second partition of unity
\be
1 = \sum_{s= 0}^{i}  v_s(r) \ ,
\ee
where
\be
\begin{cases}
v_0 &  = 1-u(M^2 r) \ , \\
v_s & = u_{s+1} \ {\rm for}\ 1\le s\le i-1 \ , \\
v_{i}(r) &  =   u(M^{2i} r) \ .
\end{cases}
\ee
Like in \cite{R1} we need $s_+ + s_- \ge i-2$ for non zero $C_{\si} (k)$, and the depth
$l(\si)$ of a sector is defined as $l = s_+ + s_-  - i+2$, with $0\le l \le i+2$. 
We have the scaled decay
(\cite{R1}, Lemma 1):

\be | \ C_{\si} (x,y) | \le c. M^{-i-l}  e^{-c' [d_{\si} (x,y)]^{\alpha}}
\ee
where $c, c'$ are some constants and
\be
d_{\si} (x,y) = M^{-i} |x_0 - y_0 |  + M^{-s_+} |x_+ - y_+ | + M^{-s_-} |x_- - y_- | \ . \label{scaleddecay}
\ee

Furthermore we recall the momentum conservation rules for the four sectors
$(\si_j)$, $j=1$, ... , $4$ meeting at any vertex 
(\cite{R1}, Lemma 4):

\medskip
\noindent{\bf Proposition 1: Momentum Conservation  at a Vertex.}
{\it The two smallest indices among $s_{j,+}$, $j=1, \ ... \ , \ 4$ differ by at most one unit,
or the smallest one, say  $s_{1,+}$ must coincide with $i_1$ with $i_1 < i_j$, $j \ne 1$.
Exactly the same statement holds independently for the minus direction.}
\medskip

We also introduce a new index for each sector, $r(\si ) = E(i(\si) +l(\si)/2)$ (where $E$ means the integer part 
like in \cite{R1}, section 4) 
and the corresponding slice propagator
\be
C_r (k) = \sum_{\si \ | \ r(\si)= r}  C_{\si } (k) \ .
\ee
We remark that this slice cutoff respects the symmetries
of the theory. It is with respect to this slice index that our main 
multislice analysis will be performed. The propagator with infrared cutoff
$r$ is defined as
\be
C_{\le r}  (k) = \sum_{\si \ | \ r(\si ) \le r }  C_{\si } (k) \ .
\ee

\section{Renormalization of the two point function}
\label{secren}

Let us define  $S_{2,\le r} (k_0, \vec k )$ as the connected two point function with 
infrared cutoff $r$, and define also~:
\begin{equation}
G_{2,\le r} (k_0, \vec k ) = \frac{1}{2} \Big(S_{2,\le r} (k_0, \vec k ) + S_{2,\le r} (-k_0, \vec k ) \Big)  \ .
\end{equation}
Consider $k$ such that $e(\vec k)=0$. If our cutoff respects the symmetries of the theory,
which is the case here, the nesting or particle-hole symmetry forces $G_2$ to vanish for such $k$. 
Using the variables $q_+$ and $q_-$ defined in (\ref {qplusorminus}), this is expressed by 

\begin{lemma}
The following equality holds~: 
\be  G_{2,\le r} (k_0 , q_+, q_- ) \Big|_{q_+ = 0 \atop {\rm or} \ q_- = 0 }  = 0 \ .
\label{belsym}
\ee
\end{lemma}

\noindent {\bf Proof~:} \ \ 
Using the symmetries of the theory, it is easy to check that for any Feynman
two point function graph $G$, the Feynman amplitude $I_G$ satisfies~:

\be
I_G(k_0, k_1 , k_2) = I_G(k_0, k_2 , k_1)  \ ,
\label{sym1}
\ee
\be
I_G(k_0, k_1 , k_2) = I_G(k_0, -k_1 , k_2) \ ,
\label{sym2}
\ee
\be
I_G(k_0, k_1 , k_2) = - I_G (-k_0, k_1 +\pi, k_2+\pi) \ .
\label{particlehole}
\ee
The last symmetry, the particle hole symmetry, is the only non-trivial
one and it can be checked because it changes
all the propagators in momentum space into their opposite with all the 
momentum conservation laws respected. Since there is an odd number of 
propagators in a two point subgraph, (\ref{particlehole}) holds.
 
Now we consider a point $\vec k$  in the first quadrant with $0\le k_1 \le \pi$ and
$0\le k_2 \le \pi$. On the Fermi 
curve whose equation in this quadrant is $k_2 = \pi - k_1$, 
we apply the relation (\ref{particlehole}) and get  
\be  0 = I_G( k_0, k_1  , k_2) + I_G(-k_0, k_1+\pi  , k_2+\pi)   = 
I_G( k_0, k_1  , k_2)  + I_G(- k_0, 2\pi -  k_2, 2\pi - k_1)\ .
\ee
By the symmetries (\ref{sym1}), (\ref{sym2}) and periodicity $2\pi$
we obtain that $I_G (k_0, \vec k)+ I_G(-k_0, \vec k) = 0$. 
By symmetry this relation holds also for 
the other quadrants, hence on all the Fermi square.

Summing over all Feynman graphs we obtain the vanishing
of $G_{2,\le r} (k_0 , q_+, q_- )$ on the Fermi surface
whose equation is  $q_+=0$ or $q_-=0$. \qed

\medskip

The function being constant on the straight lines
of the Fermi square, obviously its partial derivatives
to any order along these straight directions also vanish on the Fermi surface.

Recall that in \cite{R1} analyticity of a {\it simplified} Hubbard model at half filling
was established in a domain of the expected optimal form $|\la | \le c /|\log T|^2$. Indeed and more
precisely the result was established only for a model called "biped-free" 
in which all two point subgraphs appearing in the multislice expansion were suppressed.
A straightforward extension of the bounds given in \cite{R1} is not enough
to prove analyticity in the expected domain for the full model.

\begin{figure}[H]
\centerline{\epsfig{figure=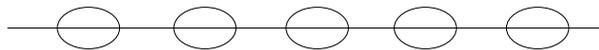,width=8cm}}
\caption{A simple chain of bipeds}
\label{figchain}
\end{figure}

Naive power counting in the style of \cite{R1} is indeed not sufficient 
to sum geometric series made of insertions of a two point subgraph at a scale
$r$ and a propagator at scale $s>>r$. Consider e.g. the simplest such sum,
made of the chain of Figure \ref{figchain}, where the three internal lines
of the biped have main scale $r$ and the external one has main scale $s>>r$. 
The naive bound for the contribution of such a chain is 
$M^{-r-l/2}$ per propagator at scale $r$, $M^{-s- l'/2}$ per propagator at scale $s$, and contains
for each irreducible biped one integral over the position of one vertex evaluated 
through the decay of a propagator of scale $s$
and one evaluated through the decay of a propagator of scale $r$. Let us neglect the auxiliary 
"depth indices" $l$ and $l'$ which are not essential.
The bound is therefore a geometric series  with ratio 
\be M^{-3r}M^{-s} M^{2r}M^{2s}= M^{s-r}\ .
\label{eqgeoratio}
\ee

This bad factor $M^{s-r}$ appears always
in the naive bounds for any similar two point function; 
it is exponential, not logarithmic in $s-r$, and certainly 
prevents a proof of analyticity, not only for $|\la | \le c /|\log T|^2$, but 
for $|\la | \le c /|\log T|^q$ for any integer $q$ as well.

As remarked in \cite{R1}, this is however only a bound, and the true contribution is much
smaller due to the particle-hole symmetry of the model at half filling.
To exploit this, and to treat the true model, we must "renormalize" 
the two point functions of the theory instead of suppressing them.
This is accomplished by a second order Taylor expansion of the
two point function with given cutoff in the style of 
\cite{DR2}.

In momentum space we change first $k_0$ to the smallest possible values
$\pm \pi T$:
\bqa  S_{2, \le r} (k_0, q_+, q_-) &=& \frac{1}{2}\Big\{ \big[S_{2, \le r} (k_0, q_+, q_-) - 
S_{2, \le r} (\pi T, q_+, q_-)\big ] \nonumber\\
&+& \big[S_{2, \le r} (k_0, q_+, q_-) - 
S_{2, \le r} (-\pi T, q_+, q_-)\big] \Big\} \nonumber\\
&+& G_{2, \le r} (\pi T, q_+, q_-) \ .
\label{momsubtrac1}
\eqa
Then we use (\ref{belsym}) to write
\bqa  G_{2, \le r} (\pi T, q_+, q_-) 
&=& G_{2, \le r} (\pi T, q_+, q_-) - G_{2, \le r} (\pi T, 0, q_-) \nonumber\\
&-& G_{2, \le r} (\pi T, q_+, 0) + G_{2, \le r} (\pi T, 0, 0) \ ,
\label{momsubtrac2}
\eqa
where the variables $(q_+, q_-)$ are the usual $k$ variables translated, so as to vanish
on the Fermi surface. They depend on the patch of coordinates chosen. This patch can be determined
by the sector of the external line to which $S_2$ is hooked.

For constructive purpose one cannot however work in momentum space
and one should write an equivalent dual formula in direct space.  In practice
a two point function $S_2$ is integrated in a bigger function against a kernel
always made of one external propagator $C$ and a rest called $R$, which 
(in momentum space) may be in general a function of the set $P_e$ of external momenta.

So in the momentum representation we have to compute not $S_2$ itself but integrals
such as
\be  I =  \int d p dq\; S_2 (p) C (q) R(p, q, P_e) 
\ee
where from momentum conservation $R(p,q, P_e) =\delta(p-q)R'(p,q, P_e) $.
To get the corresponding direct space representation we have to pass to the Fourier transform.
Using same letters for functions and their Fourier transforms we write
\be I =  \int dy d z \; S_2 (x, y ) C (y, z) R (z, x, P_e) 
\ee
(this integral being in fact by translation invariance independent of $x$) where 
\bqa S_2 (x, y ) &=& \int dp\; S_2 (p) e^{ip(x-y)}  ;  \ C (y, z) = \int dq\; C (q) e^{iq(y-z)}
; \nonumber\\
  R (z, x, P_e) &=& \int dp dq\;  R (p,q, P_e) e^{ip(z-x)} \ ,
\eqa
where the last integral is not really a double integral because of the $\delta $ function 
hidden in $R$.

Any counterterm for $I$ that is expressed in momentum space by an operator $\tau$ acting
on $S_2(p)$, such as putting $S_2$ to a fixed momentum $k$, hence $\tau S_2(p)  = S_2 (k)$,
can also be represented by a dual operator $\tau^*$ acting in direct space, but on the external
propagator. This $\tau^*$ is not unique, but a convenient choice is to use $x$ as the 
reference point for $\tau^*$:
\be  \tau I =  \int d p\; dq\; S_2 (k) C (q) R(p, q, P_e) =
\int dy\; d z\;   S_2 (x, y )   [e^{ik(x-y)} C (x, z)]  R (z, x, P_e) \ ,
\ee
hence 
\be  \tau^* C (y, z) = e^{ik(x-y)} C (x, z)\ .
\ee
The dual version of the more complicated expressions (\ref{momsubtrac1}-\ref{momsubtrac2})
is given by (we write the expressions in the patch where $q_+ = k_+-1$,  
$q_- = k_- -1$)
\bqa  I &=&  \int d p dq \ S_2 (p) C (q) R(p, q, P_e)  = I_1 + I_2
\nonumber\\ 
I_1 &=&  \int dy d z\ S_2 (x, y) \Big[ C (y , z) -\cos \big(\pi T (x_0-y_0)\big)C \big((x_0, y_+,y_-) , z\big)\Big] 
R (z, x, P_e)  \nonumber\\
I_2 &=& \int dy d z\  S_2 (x, y)  \cos \big(\pi T (x_0-y_0)\big) C \big((x_0, y_+,y_-) , z\big) R (z, x, P_e)  \nonumber\\
&-&  \int dy d z\  S_2 (x, y) \cos \big(\pi T (x_0-y_0)\big)e^{i (x_+ - y_+)} C \big((x_0, x_+,y_-) , z\big) R (z, x, P_e) 
\nonumber\\
&-&  \int dy d z\  S_2 (x, y) \cos \big(\pi T (x_0-y_0)\big)e^{i (x_- - y_-)} C \big((x_0, y_+,x_-) , z\big) R (z, x, P_e) \nonumber\\
&+&  \int dy d z\  S_2 (x, y) \cos \big(\pi T (x_0-y_0)\big)e^{i [ (x_+ - y_+)+ (x_- - y_-)]} 
C (x , z) R (z, x, P_e) 
\eqa
where the propagator $C$ is now the natural extension of the propagator to the continuum.

Each integral $I_1$ and $I_2$ will be bounded separately. We need to exploit
the differences as integrals of derivatives. This means that in $I_1$ we write~:
\bqa
C (y , z) &-& \cos \big(\pi T (x_0-y_0)\big)C \big((x_0, y_+,y_-) , z\big) \nonumber\\ 
&=& \int_0^1 dt \; \frac{d}{dt}
\Big[C \big((ty_0+(1-t)x_0, y_+,y_-) , z\big) \cos  \big(\pi T(1-t) (x_0-y_0)\big) \Big]
\nonumber\\ 
&=& \int_0^1 dt \  \frac{1}{2} (y_0-x_0) \Big[ e^{i \pi T(1-t) (x_0-y_0) } (\partial_0   + i\pi T ) 
C  \big((ty_0+(1-t)x_0, y_+,y_-) , z\big)  \nonumber\\ 
&+&  e^{-i \pi T(1-t) (x_0-y_0) } 
(\partial_0   - i\pi T ) C  \big((ty_0+(1-t)x_0, y_+,y_-) , z\big) \Big] 
\eqa
and in $I_2$ we write 
\begin{multline} C \big((x_0, y_+,y_-) , z\big) - e^{i (x_+ - y_+)} C \big((x_0, x_+,y_-) , z\big) 
- e^{i (x_- - y_-)} C \big((x_0, y_+,x_-) , z\big) \\
+ e^{i [ (x_+ - y_+)+ (x_- - y_-)]} C (x , z) = F(1,1) -  F(0,1)-  F(1,0) + F(0,0)
\end{multline} 
where
\be F(s,t) = C \Big(\big(x_0, sy_+  + (1-s)x_+, ty_-  + (1-t) y_+\big) , z\Big) 
e^{i [(1-s) (x_+ - y_+)+ (1-t)(x_- - y_-)]} \ .
\ee
Finally we can use
\be F(1,1) -  F(0,1)-  F(1,0) + F(0,0) =  \int_0^1 \int_0^1 ds dt \ \frac{d^2}{dsdt} \big(F(s,t)\big) \ .
\ee
to obtain~:
\begin{multline}
 I_2 =  \int dy d z\  S_2 (x, y) \cos \big(\pi T (x_0-y_0)\big) R (z, x, P_e) 
e^{i [(1-s) (x_+ - y_+)+ (1-t)(x_- - y_-)]} \\
\Big[  (y_+ - x_+ )(y_- - x_-  ) (\partial_+  + i )(\partial_-  + i )
C \Big(\big(x_0, sy_+  + (1-s)x_+, ty_-  + (1-t) y_+\big) , z\Big) \Big] \ . 
\label{formuleyx}
\end{multline}

\section{Multislice Expansion}

We perform a multi-slice expansion, and get a Gallavotti-Nicol\`o or clustering tree structure
as in \cite{R1}. In that paper a tree formula was used
to express a typical function for the model, namely the pressure, but the analysis applies
to any thermodynamicfunction. Now we would like to focus on the self-energy. A good starting point 
for this is the connected amputated two-point Schwinger function.

We fix here some conventions and notations that have not been introduced in \cite{R1}. We will call a "field" (between inverted commas) a five-tuple $(x, \ a, \     \sigma, \ nature, \ order)$ where~:

\be x \in V  \ , \ a \in \{ \uparrow, \downarrow \} \ , \ \sigma \in \text{Sect}(T)
\ , \ nature \ \in \{ +, \ - \} \ , \ order \ \in \{ 1, \ 2 \} .
\ee

\noindent $x$ is the spacetime position of the "field", $a$ its spin and $\sigma$ its sector. $nature$ is an element of the set whose elements are denoted $+$ and $-$; this parameter is introduced in order to distinguish between the fields and the antifields (corresponding respectively to the Grassmann variables $\psi$ and $\overline{\psi}$). Thus in the following, it may happen that we use the term field (without inverted commas) to mean a "field" such that $nature \ = +$ and of course an antifield will be a "field" such that $nature \ = -$. At last, the parameter $order$ allows to distinguish between the two copies of each field and antifield involved in the expansion of the quartic action~: $\left( \sum_{a \in \{ \uparrow, \ \downarrow \}} \overline{\psi}_a \psi_a \right) = \sum_{a, \ b} \overline{\psi}_a \psi_a \overline{\psi}_b \psi_b$, in such a way that $order \ = 1$ corresponds to the first (anti)field represented by the Grassmann variables $\overline{\psi}_a$ and $\psi_a$, while $order \ = 2$ corresponds to the second ones, represented by $\overline{\psi}_b$ and $\psi_b$. 

Given an integer $n \geq 1$, an $n$-tuple $(x_1, \ ... \ , \ x_n)$ of elements of $V$, two $n$-tuples $(a_1, \ ... \ , \ a_n)$ and $(b_1, \ ... \ , \ b_n)$ of elements of $\{ \uparrow, \ \downarrow \}$ and four $n$-tuples of elements of $\text{Sect}(T)$, denoted $(\sigma_1^j, \ ... \ , \ \sigma_n^j)$, $j \in \{ 1, \ 2, \ 3, \ 4 \}$, we define the family of the antifields~:

\be
\mathcal{AF} = \Big( (x_1, \ a_1, \ \sigma_1^1, \ -, \ 1), \ (x_1, \ b_1, \ \sigma_1^2, \ -, \ 2), \ ... \ , \ (x_n, \ a_n, \ \sigma_n^1, \ -, \ 1), \ (x_n, \ b_n, \ \sigma_n^2, \ -, \ 2)    \Big)\ .
\ee

We can imagine it as a $2n$-tuple indexed by the set $[n] \times \{ 1, \ 2 \}$ (where $[n]$ denotes the set $\{ 1, \ ... \ , \ n  \}$), lexicographically ordered~:
\be
(1,\ 1) \prec (1, \ 2) \prec (2, \ 1) \prec (2, \ 2) \prec \ ... \ \prec (n, \ 1) \prec (n, \ 2) \ .
\ee
In the same way we introduce the family of the fields~:
\be
\mathcal{F} = \Big( (x_1, \ a_1, \ \sigma_1^3, \ +, \ 1), \ (x_1, \ b_1, \ \sigma_1^4, \ +, \ 2), \ ... \ , \ (x_n, \ a_n, \ \sigma_n^3, \ +, \ 1), \ (x_n, \ b_n, \ \sigma_n^4, \ +, \ 2)    \Big)\ .
\ee
Observe that $\mathcal{AF}$ and $\mathcal{F}$ are defined as families and not as sets. Hence the cardinality of $\mathcal{AF}$ and $\mathcal{F}$ is $2n$, whatever may be the values of the parameters $\{ x_v \}$, $\{ a_v \}$, $\{ b_v \}$ and $\{ \sigma_v^j \}$.

Given $f \in \mathcal{AF}$ and $g \in \mathcal{F}$, we will simply denote by $C(f,g)$ the propagator~:
\be
C(f,g) = \delta_{a(f), \ a(g)} \delta_{\sigma(f), \ \sigma(g)} C \Big(x(f) - x(g)\Big) \ ,
\ee
where the notations $a(f), \ a(g), \ \sigma(f), \ \sigma(g), \ x(f), \ x(g)$ have an immediate obvious meaning. 

With all these notations, we can express the partition function of the model as~:
\be
Z(V) = \sum_{n = 0}^{\infty} \frac{\lambda^n}{n!} \int_{V^n} d^3x_1 \ ... \ d^3x_n \sum_{\{ a_v \}, \ \{ b_v \}} \sum_{\{ \sigma_v^j \}} \det_{(f,g) \in \mathcal{AF} \times \mathcal{F}} \Big( C(f,g) \Big) \ .
\ee
Sometimes we shall write simply $\left\{ \begin{matrix} \mathcal{AF}  \\  \mathcal{F} \end{matrix} \right\}  $ for the Fermionic determinant (Cayley's notation). To write the unnormalized unamputated two-point Schwinger function~:
\be
S_2 (Y, \ Z)_{\sigma_0} = \int d\mu_C  \ (\overline{\psi}, \ \psi) \ \overline{\psi}_{\uparrow, \ \sigma_0} (Y) \psi_{\uparrow, \ \sigma_0} (Z) \exp \left( \lambda \int_{V} d^3x  \ \Big( \sum_{a} \overline{\psi}_a (x) \psi_a(x) \Big)^2 \right) \ ,
\ee
we only need to add the source terms $(Y, \ \uparrow, \ \sigma_0, \ -)$ to $\mathcal{AF}$ and $(Z, \ \uparrow, \ \sigma_0, \ +)$ to $\mathcal{F}$\footnote{Note that these two external "fields" have no $order$ parameter.}. 
Since $\mathcal{AF}$ and $\mathcal{A}$ are indeed totally ordered families, we must specify in which position $(y, \ \uparrow, \ \sigma_0, \ -)$ and $(z, \ \uparrow, \ \sigma_0, \ +)$ are inserted. Clearly, they must be added in first position, 
that is, we have~:
\be
\mathcal{AF} = \Big( (y, \ \uparrow, \ \sigma_0, \ -), \ (x_1, \ a_1, \ \sigma_1^1, \ -, \ 1), 
\ ... \ , \ (x_n, \ a_n, \ \sigma_n^1, \ -, \ 1), \ (x_n, \ b_n, \ \sigma_n^2, \ -, \ 2)    \Big)
\ee
and
\be
\mathcal{F} = \Big( (z, \ \uparrow, \ \sigma_0, \ +), \ (x_1, \ a_1, \ \sigma_1^1, \ -, \ 1), 
\ ... \ , \ (x_n, \ a_n, \ \sigma_n^1, \ -, \ 1), \ (x_n, \ b_n, \ \sigma_n^2, \ -, \ 2)    \Big)\ .
\ee
Observe that, with a slight abuse of notation, we denote these two families again by $\mathcal{AF}$ and $\mathcal{F}$. With this convention, the expression of
the two point function $S_2 (y, \ z)_{\sigma_0}$ is exactly the same as the one of $Z(V)$~:

\be
S_2 (Y, \ Z)_{\sigma_0} = \sum_{n = 0}^{\infty} \frac{\lambda^n}{n!} \int_{V^n} d^3x_1 \ ... \ d^3x_n \sum_{\{ a_v \}, \ \{ b_v \}} \sum_{\{ \sigma_v^j \}} \det_{(f,g) \in \mathcal{AF} \times \mathcal{F}} \Big( C(f,g) \Big) \ .
\ee
The main tool to express the {\it connected} two point function is a Taylor jungle formula \cite{AR}, that is a forest formula which is ordered according to the main slice index namely $r$ attached to the propagator, to expand the Fermionic determinant. To extract the {\it connected} part of the two-point function, namely $S_2 (Y, \ Z)_{c, \ \sigma_0}= Z^{-1} S_2 (Y, \ Z)_{\sigma_0}$,  we only need to factorize the contributions of the vacuum clusters of the jungle, and we get a {\it tree formula}~:

\begin{multline}
S_2 (Y, \ Z)_{c, \ \sigma_0} = \sum_{n = 0}^\infty \frac{\lambda^n}{n!} \int_{V^n} d^3x_1 \ ... \ d^3x_n \sum_{\{ a_v \}, \ \{ b_v \}, \ \{ \sigma_v^j \}} \sum_{\text{oriented trees} \mathcal{T} \atop \text{over} \ \mathcal{V}} \sum_{\text{field attributions} \atop \Omega} \\
\left( \prod_{\ell \in \mathcal{P}_2(\mathcal{V})}  \int_0^1 dw_{\ell}  \right)
\left( \prod_{\ell \in \mathcal{T}} C \Big( f(\ell, \Omega), \ g(\ell, \Omega) \Big) \right) \det_{(f, \ g) \in \mathcal{AF}_{\text{left}} \times \mathcal{F}_{\text{left}}} \Big( C(f, \ g, \{ w_\ell \})  \Big) \ .
\end{multline}

The amputated connected two point function $S_2 (y, \ z)_{c, a}$ is given by a similar
formula, in which we should delete the two external sources $Y$ and $Z$ and the two propagators which connect them to two particular external distinguished vertices\footnote{Indeed we can forget the graphs where these two external sources $Y$ and $Z$ connect to the same external vertex, the "generalized tadpoles", since they are zero by the particle hole symmetry.}.
Let us rename the position of these vertices as $y$ and $z$,
and rename all remaining internal positions as $x_1 , ..., x_n$.
So, after integration over positions of these $n$ internal vertices, this amputated function 
is a function of the positions $y$ and $z$ of the two particular special external vertices.

We shall denote $\mathcal{V}$ the family of the vertices~: $\mathcal{V} = (y, \ z, \ x_1, \ ... \ , \ x_n)$.

We recall that a tree over $\mathcal{V} = \{ y, \ z, \ x_1, \ ... \ , \ x_n \}$ is a set of pairs of vertices $\{v, \ v'  \}$ (called the links of the tree), such that the corresponding graph has no loop and connects all the elements of $\mathcal{V}$. As $|\mathcal{V}| = n+2$; any tree over $\mathcal{V}$ has $n+1$ links.

Once a tree $\mathcal{T}$ over $\mathcal{V}$ is chosen, a field attribution $\Omega$ for $\mathcal{T}$ is a family of the form \\ $\Big( (\omega_\ell, {\omega'}_\ell ) \Big)_{\ell \in \mathcal{T}}$ where $\omega_\ell$ is a map from the pair $\ell$ to $\{ 1, \ 2 \}$ and ${\omega'}_\ell$ a one-to-one map from $\ell$ to $\{ +, \ - \}$. Hence $\Omega$ is simply the choice, for each "half-line" of the tree $\mathcal{T}$ of a precise "field" of the vertex to which this half-line hooks. We have taken into account the constraint that a field must contract with an antifield by the fact that the maps $\omega_\ell \ \colon \ \ell \mapsto \{ +, \ - \}$ are one-to-one.

Given $\ell \in \mathcal{T}$ and a field attribution $\Omega$, we denote respectively by $f(\ell, \ \Omega)$ and $g(\ell, \ \Omega)$ the antifield and the field attached to $\ell$ by $\Omega$. $\mathcal{AF}_{\text{left}}$ and $\mathcal{F}_{\text{left}}$ are the families of the remaining "fields"~:
\begin{equation}
\mathcal{AF}_{\text{left}} = \mathcal{AF} \backslash \{ f(\ell, \ \Omega), \ \ell \in \mathcal{T} \} \ \text{and} \ \mathcal{F}_{\text{left}} = \mathcal{F} \backslash \{ g(\ell, \ \Omega), \ \ell \in \mathcal{T} \} \ .
\end{equation}

At last we must precise the expression of the entries of the remaining Fermionic determinant that depends now on the interpolation parameters $\Big( w_\ell \Big)_{\ell \in \mathcal{P}_2(\mathcal{V})}$. We recall that (see \cite{AR})-(\cite{R1} for details) the data $w_\ell$ allows to define a vector $X^{\mathcal{T}} \Big( \{ w_\ell \} \Big)$ whose components are indexed by $\mathcal{P}_2(\mathcal{V})$, the set of the (unordered) pairs of vertices. By definition, for $\{ v, \ v' \} \in \mathcal{P}_2(\mathcal{V})$, $X^{\mathcal{T}} \Big( \{ w_\ell \} \Big)_{\{ v, \ v' \}}$ is the infimum of the $w_\ell$ parameters over the unique path in $\mathcal{T}$ from $v$ to $v'$. Then, the expression of $C \Big( f, \ g, \ \{ w_\ell \} \Big)$ is simply obtained by multiplying $C(f, \ g)$ by the component of $X^{\mathcal{T}} \Big( \{ w_\ell \} \Big)$ corresponding to the vertices $v(f)$ and $v(g)$ of $f$ and $g$. Hence we have~:
\begin{equation}
C \Big( f, \ g, \ \{ w_\ell \} \Big) = X^{\mathcal{T}} \Big( \{ w_\ell \} \Big)_{\{ v(f), \ v(g) \}} C(f,g) \ .
\end{equation}

\subsection{The Gallavotti-Nicol\`o tree}

In order to analyze further this sum, it is well known that the main tool is the
"Gallavotti-Nicol\`o" or clustering tree which represents the inclusion relations of the connected
components of "higher scales" (smaller $r$ indices) into those of  "lower scales" 
(bigger $r$ indices) \cite{BG}. This tree is also the key tool to identify the components that require
some renormalization (here the two point functions). 
But before doing this, we want to describe precisely the constraints on the sum over the sectors $\{ \sigma_v^j \}$. Indeed, this sum could be let free of constraints, but due to the expression of the propagator~:
\be
C (f, \ g) = C \big( (x(f), \ a(f), \ \sigma(f)); \ (x(g), \ a(g), \ \sigma(g) ) \big) = \delta_{a(f), \ a(g)} \delta_{\sigma(f), \ \sigma(g)} C_{\sigma(f)} (x(f), \ x(g)) \ ,
\ee
we see easily that the sectors and spin indices are conserved along each line of the tree $\mathcal{T}$. Therefore, once $\mathcal{T}$ has been fixed, the sum over the $\sigma_v^j$'s can be understood as a sum over the families of sectors indexed by the lines of $\mathcal{T}$, denoted $\big( \sigma_\ell \big)_{\ell \in \mathcal{T}}$, and the families of sectors indexed by the remaining "fields", $\big( \sigma_f \big)_{f \in \mathcal{AF}_{\text{left}} \cup \mathcal{F}_{\text{left}}}$.

Now let us suppose we are given an oriented tree $\mathcal{T}$ over $\mathcal{V}$, and an attribution of sectors, $\big( \sigma_\ell \big)_{\ell \in \mathcal{T}}$ and $\big( \sigma_f \big)_{f \in \mathcal{AF}_{\text{left}} \cup \mathcal{F}_{\text{left}}}$. 
The Gallavotti-Nicol\`o tree is defined as follows~: 
for each index $r \in [0, \ r_{max}(T)]$, we define a partition $\Pi_r$. $\Pi_r$ is the set of the connected components of the graph whose set of vertices is $\mathcal{V}$ and whose internal
tree lines are the lines of $\mathcal{T}$ such that $r_\ell \leq r$. The family $\bigcup_{r \in [0, \ r_{max}]} \Pi_r$ is partially ordered by the inclusion relation and forms the nodes of the Gallavotti-Nicol\`o tree.

\begin{figure}[H]
\centerline{\psfig{figure=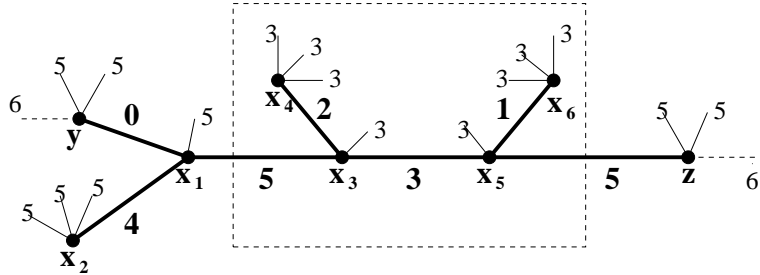,width=10cm}}
\caption{A contribution with eight vertices to the two point function at scale 6}
\label{sampletree}
\end{figure}

To visualize better the situation, let us take the example of Figure \ref{sampletree}
for an amputated two point function with external
vertices at $y$ and $z$ (the external amputated legs in slice 6 are represented as dotted lines
in Figure \ref{sampletree}). The total number of vertices is 8,
hence there are 7 lines in the tree $\mathcal{T}$ represented as bold lines, and 16 internal fields
in the determinant represented as thin half-lines.
 
In the attribution of $r$ indices chosen we see that there is a two point subfunction to renormalize,
the one in the dotted box, which is completed at scale 3 with external lines at scale 5.

\begin{figure}[H]
\centerline{\psfig{figure=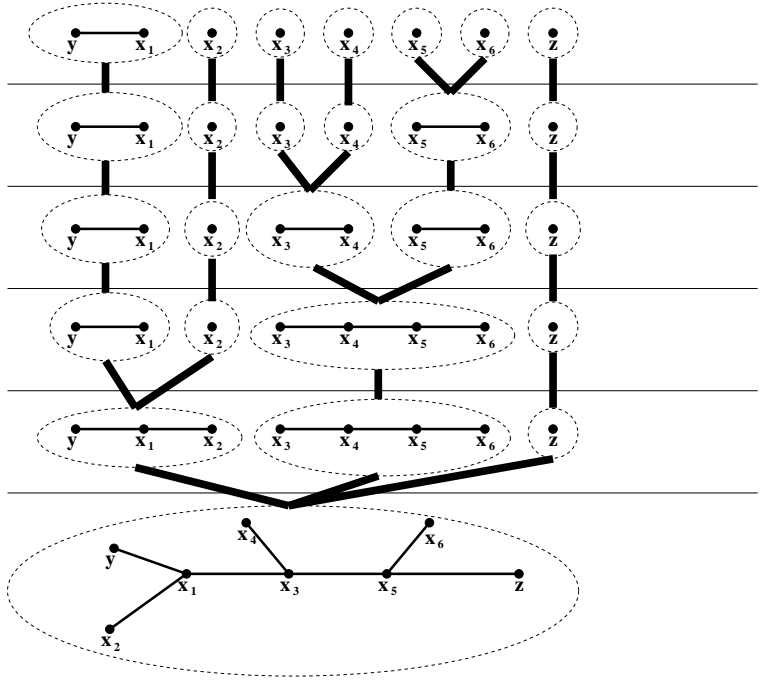,width=10cm}}
\caption{The Gallavotti-Nicol\`o tree corresponding to Figure \ref{sampletree}}
\label{gallani}
\end{figure}

The corresponding Gallavotti-Nicol\`o tree is pictured in Figure \ref{gallani} (with determinant fields omitted for simplicity). This abstract tree should not be confused with $\mathcal {T}$: its lines are the
bold lines of Figure \ref{gallani}.

As in \cite{R1}, we can now write an expression of $S_2 (y,z)_{c, a}$ re-ordered in terms of these "clustering tree structures", in which all nested sums have to be compatible~: 

\begin{multline}
S_2 (y, \ z)_{c, a} = \sum_{n = 0}^\infty \frac{\lambda^{n+2}}{n!} \int_{V^n} d^3x_1 \ ... \ d^3x_n \sum_{\{ a_v \}, \ \{ b_v \}} \sum_{\text{clustering tree} \atop \text{structures} \ \mathcal{C}} \sum_{\text{trees} \mathcal{T} \atop \text{over} \ \mathcal{V}} \sum_{\text{field attributions} \atop \Omega} \sum_{\{ \sigma_v^j \}}\\
\left( \prod_{\ell \in \mathcal{P}_2(\mathcal{V})}  \int_0^1 dw_{\ell}  \right)
\left( \prod_{\ell \in \mathcal{T}} C \Big( f(\ell, \Omega), \ g(\ell, \Omega) \Big) \right) \det_{(f, \ g) \in \mathcal{AF}_{\text{left}} \times \mathcal{F}_{\text{left}}} \Big( C(f, \ g, \{ w_\ell \})  \Big) \ .
\end{multline}

In the Gallavotti-Nicol\`o tree, of particular interest to us are the nodes 
such as the dotted box of Figure \ref{sampletree} between scales 3 and 5
which correspond to {\it two}
point functions. They are the ones that were artificially suppressed in the simplified model \cite{R1}.
We need to renormalize them to solve the divergent power counting explained in section \ref{secren}.
But we can choose to renormalize only the two point functions for which external lines
have $r$ index bigger than the maximum index of internal lines plus 2,
so as to create a gap betwen internal and external 
supports\footnote{The two point functions for which the external 
$r$ index is the maximum $r$ index of internal lines plus 1 don't really need
renormalization, as is obvious from power counting (see (\ref{eqgeoratio})).}. Such two point functions
are the {\it dangerous} nodes of the GN tree.
The gap ensures that all such {\it dangerous}
two point functions, which are those that we need to renormalize,
are automatically one-particle irreducible by momentum conservation\footnote{Indeed any one particle reducible two point function would have its external momentum also flowing through
any internal one-particle-reducibility line, which is a contradiction with the fact that 
the internal and external cutoffs have empty intersection.}.
Hence they correspond to the so-called self-energy. 

We can re-order the expression of $S_2(y, \ z)_{c, a}$ in terms of these dangerous two point subgraphs, in the spirit of \cite{R1}~: 

\begin{multline}
S_2 (y, \ z)_{c, \ a} = \sum_{n = 0}^\infty \frac{\lambda^{n+2}}{n!} \int_{V^n} d^3x_1 \ ... \ d^3x_n \sum_{\{ a_v \}, \ \{ b_v \}}
\\
 \sum_{\text{biped structures} \atop \mathcal{B}} \sum_{\text{external fields} \atop \mathcal{EB}} \sum_{\text{clustering tree} \atop \text{structures} \ \mathcal{C}} \sum_{\text{trees} \mathcal{T} \atop \text{over} \ \mathcal{V}} \sum_{\text{field attributions} \atop \Omega} \sum_{\{ \sigma_v^j \}}\\
\left( \prod_{\ell \in \mathcal{P}_2(\mathcal{V})}  \int_0^1 dw_{\ell}  \right)
\left( \prod_{\ell \in \mathcal{T}} C \Big( f(\ell, \Omega), \ g(\ell, \Omega) \Big) \right) \det_{(f, \ g) \in \mathcal{AF}_{\text{left}} \times \mathcal{F}_{\text{left}}} \Big( C(f, \ g, \{ w_\ell \})  \Big) \ .
\label{sommecompliq}
\end{multline}

\section{Main theorem on the self-energy}
\label{maintheo}

We have given in the last section an expression for the connected amputated 2-point Schwinger function. Now we would like to consider the self-energy $\Sigma(y,z)$.
This quantity can be defined either through its Feynman graph expansion,
or through a Legendre transform.

In the first approach, which we use, $\Sigma(y,z)$ is given by the same sum  (\ref{sommecompliq})
than $S_2 (y, \ z)_{c, \ a} $ but restricted to the
contributions which are 1-particle-irreducible in the channel $y-z$,
that is, in which $y$ and $z$ cannot be disconnected by the deletion of a single line. 
This definition does not look very
constructive because in principle we would have to expand out all the remaining determinant in   (\ref{sommecompliq}) to know which contributions are 1-PI or not. But in the next section
we shall see that to extract this information a partial (still constructive) expansion 
of the determinant is enough.

In this section we only formulate our main bound on this connected amputated and 
one particle irreducible (1-PI) 2-point function or self-energy $\Sigma$. 
Note that, for convenience, we shall simply write in the following "1-PI" 
to mean~:"1-particle-irreducibility in the channel $y-z$".

The sum of all contributions
to the self-energy with infrared cutoff  $r$ and fixed external positions
$y$ and $z$ will be called $\Sigma_{2}(y,z)^{\le r}$.

Consider the set $\Si_r$
of triplets $\bar \si = (i (\bar \si), s_{+}(\bar \si) , s_{-}(\bar \si))$ 
with $0\le i \le r$ and $0\le s_{\pm} \le r$, also called "generalized sectors".
We can obviously also define the scale distance $d_{ \bar \si }(y,z)$ for such triplets as in (\ref{scaleddecay}),
and the index $r(\bar \si) =(i (\bar \si) + s_{+}(\bar \si) + s_{-}(\bar \si))/2 $ . Then 
with all the notations of the previous section, the following
bound holds~:

\begin{thm} \label{theortwopoint}There exists a constant $K$
such that~:
\be
|\Sigma_{2}(y,z)^{\le r}| \leq  (\lambda | \log T |)^2 \sup_{\bar \si \in \Si_r }
K M^{-3r(\bar\si) } e^{ -cd_{ \bar \si }^{\alpha}(y,z) }  \label{eqimporta}
\ee
\be
|y_+ - z_+|.|y_- - z_-| .|\Sigma_{2}(y,z)^{\le r}| \leq  (\lambda | \log T |)^2 \sup_{\bar \si \in \Si_r}
K M^{-2r(\bar\si) } e^{ -cd_{ \bar \si }^{\alpha}(y,z) }  \label{eqimportant}
\ee
\be
|y_0 - z_0| .|\Sigma_{2}(y,z)^{i_0}| \leq  (\lambda | \log T |)^2 \sup_{\bar \si \in \Si_r}
K M^{-2r(\bar\si)  } e^{ -cd_{ \bar \si }^{\alpha}(y,z) }  \label{eqimportant0}
\ee
\end{thm}

For the second equation  (\ref{eqimportant}), a naive bound would have $M^{-r}$ instead
of $M^{-2r}$.
So the crucial point is to gain a factor $M^{-r}$ in the
bound (\ref{eqimportant}). (\ref{eqimporta}) and (\ref{eqimportant0}) are easy.

The next four sections are dedicated to the proof of this theorem. 
We call a self energy contribution "primitively divergent"
if there is no smaller biped in it. The sum of all such "primitively divergent" contributions
to the self-energy with infrared cutoff  $r$ and fixed external positions
$y$ and $z$ is called $\Sigma_{2,pr}(y,z)^{\le r}$. We first prove in the next three sections
that the bounds (\ref{eqimportant}) and (\ref{eqimportant0}) hold for $\Sigma_{2,pr}(y,z)^{\le r}$,
then by an inductive argument we extend the bound to the general unrestricted self-energy. 

The most naive bounds don't work. Indeed we should optimize power counting and positions integrals separately 
in the 0 and $\pm$ directions  in order to bound correctly the effect
of the $(y-z)_{\pm}$ factors in (\ref{eqimportant}). But the problem is how to do this
constructively. One cannot simultaneously build the three spanning trees that would optimize spatial integrations with respect to the 0 and 
$\pm$ directions, as this may typically develop too many loops out of the determinant.  
The road to solve this problem is to derive
not only a 1-PI, but a  2-PI expansion inside each two point contribution to renormalize.
This expansion can be controlled constructively;
then one can optimize over the 0 and $\pm$ multiscale analysis, 
using only the tree $\mathcal{T}$ and the additional loops which 
the expansion has taken out of the determinant.

In this way one obtains a better bound
 than the one obtained naively by simply exploiting a single tree formula as in
\cite{R1}. This is the key to our problem of the
renormalization of the 2-point function.

\section{Multiarch expansion}

Consider the self-energy of the model. 
The previous tree expansion insured the connexity of the graphs but not their 1 or 2-particle-irreducibility. 
We are going now to expand out explicitly some additional lines from the determinant, in order to complete the tree $\mathcal{T}$ into a 2-PI graph. Nevertheless it is not trivial to insure that this additional expansion does not generate "too many" terms, or in other words that it is "constructive". In the following section, we explain in detail this expansion for an expression of the type $F=\prod_{\ell \in \mathcal{T}} C_{\sigma(\ell)} (f(\ell),g(\ell)) \det_{\text{left}, \mathcal{T}}$.

\subsection{1-particle-irreducible arch expansion}

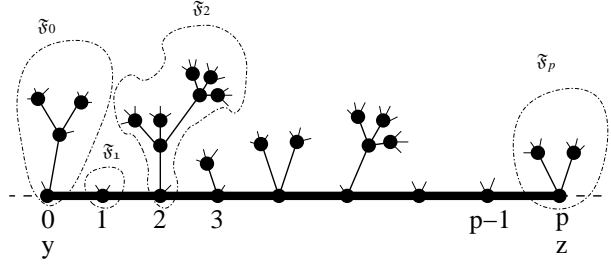
\begin{figure}[H]
\centerline{\resizebox{8cm}{!}{\input{arbre.pstex_t}}}
\caption{The tree $\mathcal{T}$ and the "field packets" $\mathfrak{F}_0$, ..., $\mathfrak{F}_p$.}
\label{tree}
\end{figure}

First, we fix some conventions. We consider the tree $\mathcal{T}$ connecting all the vertices $y, \ z,$ $x_1, \ ... \ , \ x_n$. We distinguish in $\mathcal{T}$ the unique path connecting $y$ and $z$ through $\mathcal{T}$, denoted by $P(y, \ z, \ \mathcal{T})$. Each vertex of this path is numbered by an integer starting with 0 for $y$ and increasing towards $z$, which is the end of the path (with number $p$). The set of the remaining $2(n+2)$ fields and antifields, denoted by $\mathfrak{F}_{\text{left},\mathcal{T}} = \mathcal{AF}_{\text{left},\mathcal{T}} \cup \mathcal{F}_{\text{left},\mathcal{T}}$, is divided into $p +1$ disjoint subsets or "packets" $\mathfrak{F}_0, \ ... \ , \ \mathfrak{F}_p$~: by definition, an element $f \in \mathfrak{F}_{\text{left},\mathcal{T}}$ belongs to $\mathfrak{F}_k$ if and only if $k$ is the first vertex of $P(y, \ z, \ \mathcal{T})$ met by the unique path in $\mathcal{T}$ joining the vertex to which $f$ is hooked to $y$. Figure \ref{tree} allows to visualize better the situation. When $f$ belongs to the packet  $\mathfrak{F}_k$ we also say that the packet index of $f$ is $k$.

In this figure we have represented the external (amputated) propagators by dotted lines, the links of $P(y,z,\mathcal{T})$ by bold lines, the other links of $\mathcal{T}$ by thin lines and at last the remaining fields in the determinant by thinner half-lines.

Once the ordered family of subsets of fields $\mathfrak{F}_0$, ..., $\mathfrak{F}_p$ has been defined, the arch expansion is carried out in the standard way of \cite{DR2}, Appendix B1. 

\begin{figure}[H]
\centerline{\resizebox{8cm}{!}{\input{arbre2.pstex_t}}}
\caption{The tree $\mathcal{T}$ completed by a first line from the arch expansion}
\label{tree_one_arch}
\end{figure}
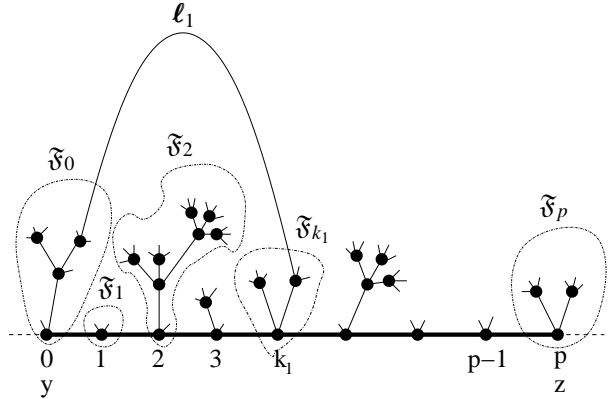

Let us recall this expansion here for self-completeness.
Among all the possible contraction schemes implicitly contained in $\det_{\text{left}, \ \mathcal{T}}$, we select through a Taylor expansion step with an interpolating parameter $s_1$ those which have a contraction between an element of $\mathfrak{F}_0$ and $\cup_{k = 1}^{p} \mathfrak{F}_k$. Given such a contraction, we call $k_1$ the index of the precise packet joined to $\mathfrak{F}_0$ by this contraction. Thus we have added to $\mathcal{T}$ an explicit line $\pmb{\ell}_1$ joining $\mathfrak{F}_0$ to $\mathfrak{F}_{k_1}$. At this stage, the graph obtained is 1-particle-irreducible in the channel $y - x_{k_1}$ (see Figure \ref{tree_one_arch}).

Then we continue the procedure, testing whether there is a contraction between an element of
$\cup_{k = 0}^{k_1} \mathfrak{F}_k$ and one of $\cup_{k = k_1 + 1}^{p} \mathfrak{F}_k$. If there is not, the line from $k_1$ to $k_1 + 1$ of the path $P(x, \ y, \  \mathcal{T})$ is certainly a line of 1-particle-reducibility (i.e. its deletion would disconnect $y$ and $z$), and therefore the corresponding contraction schemes  do not contribute to the self-energy.
But on the contrary, if there exists a line $\pmb{\ell}_2$ between $\cup_{k = 0}^{k_1} \mathfrak{F}_k$ and $\cup_{k = k_1 + 1}^{p} \mathfrak{F}_k$, we select it and we have the picture of Figure \ref{tree_two_arches}:

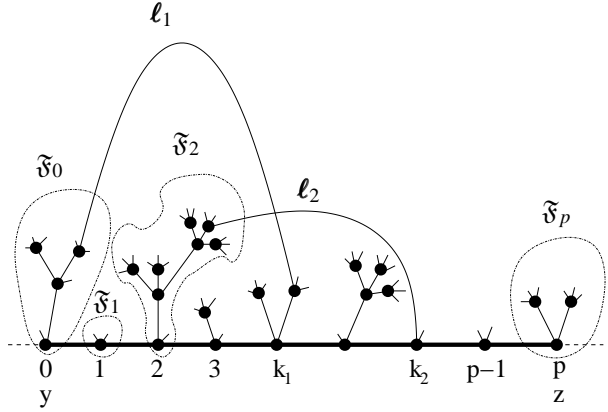
\begin{figure}[H]
\centerline{\resizebox{8cm}{!}{\input{arbre3.pstex_t}}}
\caption{The tree $\mathcal{T}$ completed by two lines from the arch expansion}
\label{tree_two_arches}
\end{figure}

The graph $\mathcal{T} \cup \{ \pmb{\ell}_1, \ \pmb{\ell}_2  \}$ is clearly 1-particle-irreducible in the channel $y - x_{k_2}$. Observe that $0 < k_1 < k_2$ therefore, in at most $p$ steps, we shall reach certainly the end vertex $z$ and we shall have a 1-particle-irreducible graph (in the channel $y - z$).  
Any final set of $m$ arches derived in this way is called an $m$-arch system.
We obtain the 1-PI part of the determinant as~:
\begin{multline}
F_{1-\text{PI}} = \sum_{{
m-{\rm arch\  systems} \atop
\big( (f_1,g_1),...,(f_m,g_m)\big) }\atop
{\rm with}\  m \leq p}
\left[ \prod_{r = 1}^{m} \int_0^1 ds_r \right] \left( \prod_{r = 1}^m C(f_r,g_r) (s_1, \ ... \ , s_{r-1}) \right) \frac{\partial^m \det_{\text{left}, \ \mathcal{T}}}{\prod_{r = 1}^m \partial C(f_r,g_r)} \Big( \{ s_r \} \Big) \ .
\end{multline}

The expansion respects positivity of the interpolated propagator at any stage,
because all $s_r$ interpolations are always performed between a subset of packets and 
its complement, hence the final covariance as function of the $s_r$ parameters
is a convex combination with positive coefficients of block-diagonal covariances. This ensures
that the presence of the $s_r$ parameters does not alter Gram's bound
on the remaining determinant, which is the same than with all these 
parameters set to 1 (\cite{DR1}-\cite{DR2}).

Furthermore it is constructive in the sense that it does not generate any factorial
in the bounds for the sum over all derived arches. Here is a subtlety. Once the departure and arrival fields joined by the arches have been fixed (which costs at most $4^n$), the arrival fields are determined
because their packet indices are strictly growing. But the departure fields are not, and in principle this could create a constructive problem. 

\begin{figure}[H]
\centerline{\resizebox{8cm}{!}{\input{arbre4.pstex_t}}}
\caption{A pair of arches which is not minimal from the 1-PI point of view}
\label{two_bad_arches}
\end{figure}
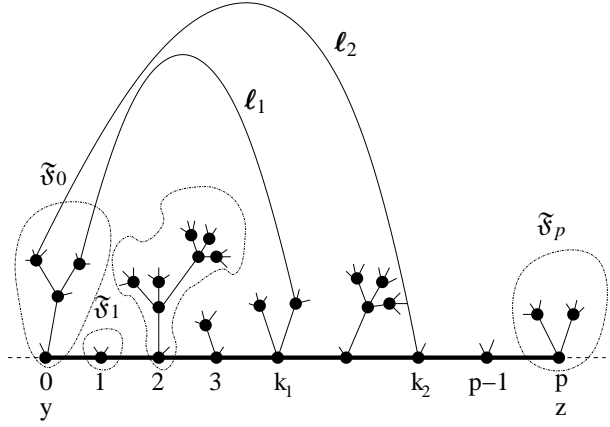

For example, if the line $\pmb{\ell}_1$ joins $\mathfrak{F}_0$ to $\mathfrak{F}_{k_1}$, it is possible for the second one, $\pmb{\ell}_2$, to join $\mathfrak{F}_0$ to $\mathfrak{F}_{k_2}$ (see figure \ref{two_bad_arches}). Remark that in this case {\it a posteriori} ${\pmb \ell}_1$ is useless.

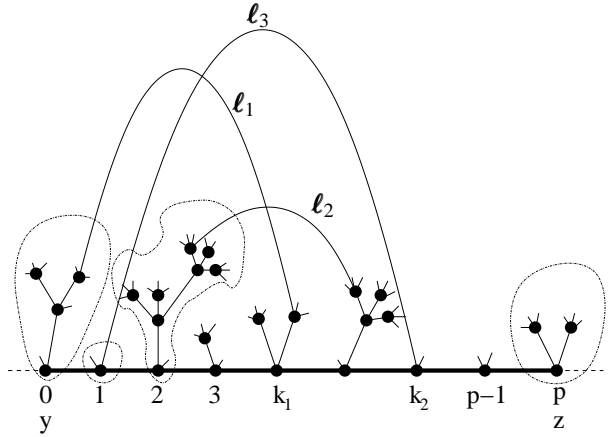
\begin{figure}[H]
\centerline{\resizebox{8cm}{!}{\input{arbre5.pstex_t}}}
\caption{Another example of a "non-minimal" system of three arches}
\label{three_bad_arches}
\end{figure}

With three arches, an arch system such as Figure \ref{three_bad_arches} shows the same phenomenon, in the sense that  {\it a posteriori} $\pmb{\ell}_2$ is useless. 

This is not a great disadvantage, because in spite of this lack of minimality, 
the expansion can indeed be controlled in a constructive way.
The reason is that the arches for which the departure fields indices do not grow are damped 
by small $s$ interpolation parameters, so that the result is indeed bounded
by $K^n$ \cite{DR2}. More precisely the dependence in the $s_r$ parameters 
in front of each arch system is a monomial $\prod_{r = 1}^m s_r^{q_{r, m-{\rm arch}}}$ , so 
that we have:
\be 
\left( \prod_{r = 1}^m C(f_r,g_r) (s_1, \ ... \ , s_{r-1}) \right)
= \prod_{r = 1}^m C(f_r,g_r)  \prod_{r = 1}^m s_r^{q_{r, m-{\rm arch}}} \ .
\ee
The reader can check that the integer 
$q_{r, m-{\rm arch}} \ge 0$ is the number of arches which fly entirely over the $r$-th arch,
making it useless.

\begin{lemma}
There exists some numerical constant $K$ such that (for $n \ge 1$)~:
\be
\sum_{m=1}^{p}
\sum_{{m-{\rm arch\ systems} \atop \big( (f_1,g_1), \ ... \ , \ (f_m,g_m) \big)} \atop {\rm with} \ m \leq p} 
\left( \prod_{r=1}^m \int_0^1 ds_r \right)  \prod_{r=1}^m  s_r^{q_{r, m-{\rm arch}}}   \le c.K^n\ .
\ee  \label{lemmeconstr}
\end{lemma}

\noindent{\bf Proof} 
The proof is identical to \cite{DR2}, Lemma 9. We reproduce it here for completeness.
Consider ${\mathfrak{F}}_{k_r}$  the arrival packet of the $r$-th arch,
which joins the field $f_r $ to the field $g_r \in {\mathfrak{F}}_{k_r}$. The set of possible departure packets
to which $f_r$ must belong is
\be  {\mathfrak{E}}_r = {\mathfrak{F}}_0 \cup {\mathfrak{F}}_1 \cup ... \cup {\mathfrak{F}}_{k_{r-1}}
\ee
We also define $e_i = \vert {\mathfrak{E}}_i - {\mathfrak{E}}_{i-1} \vert $ as the number of fields
and antifields in ${\mathfrak{E}}_i$ and not in ${\mathfrak{E}}_{i-1}$.

The sum over all $m$-arch systems which we have to bound is
\be
\sum_{m=1}^{p}\; \sum_{0<k_{1}<...<k_{m}=p}\;
\sum_{g_r \in {\mathfrak{F}}_{k_r}\atop r=1,...,m} 
\int_0^1 ds_1...\int_0^1 ds_m\;  
\sum_{f_r \in {\mathfrak{E}}_{r}\atop r=1,...,m}  \prod_{r=1}^m  s_r^{q_{r, m-{\rm arch}}} \ .
\ee
We start observing that
\be
\sum_{f_r \in {\mathfrak{E}}_{r}\atop r=1,...,m}   \prod_{r=1}^m  s_r^{q_{r, m-{\rm arch}}}   \le
 \prod_{r=1}^{m} a_{r}(s_{1},...,s_{r-1})  \label{sestimate}
\ee
where $a_r$ is defined inductively by
$a_{1}= e_{1} $ and
$a_{r}(s_{1},...,s_{r-1}) = e_{r}+s_{r-1} a_{r-1}(s_{1},...,s_{r-2})$.
To see this we remark that  we have $e_1$ choices
to choose $f_1$.  In the same way, we have $e_2$ choices to choose
$f_2$  if it does not hook to ${\mathfrak{F}}_1$.  
If it does hook to ${\mathfrak{F}}_1$, we have $e_1=a_1$  choices, but we also have a multiplicative 
factor $s_1$ coming from $s_1^{q_{1, m-{\rm arch}}}$ . This iterates into (\ref{sestimate}). Remark that (\ref{sestimate})
is an overestimate, not an equality, as, once $f_1$ is fixed 
we have only $e_1-1$ choices for $f_2$ if it falls in the ${\mathfrak{F}}_1$ packet, and so on.

We have also
\be
\int_{0}^{1} \prod_{r=1}^{m} ds_{r} \prod_{r=1}^{m} 
a_{r}(s_{1},...,s_{r-1})  \le e^{\sum_{r=1}^{q} e_{r}}.
\ee
Indeed this follows from the inductive use of  
\be
\int_{0}^{1} (as +b ) ds \le \int_{0}^{1} e^{as +b } ds  \le (1/a)e^{a +b}\; , \ {\rm for} \  a>0, b>0.
\ee
Now, as  $e_{r}= \vert {\mathfrak{E}}_r\backslash {\mathfrak{E}}_{r-1} \vert $, we have
\be
\sum_{r=1}^{m} e_{r}\le  2 (n +2)
\ee
since $2(n+2)$ is the total number of remaining fields (after extraction of the tree) in the amputated
two point function considered.

Finally it is easy to check that
\be
\sum_{m} \;\sum_{0<k_{1}<...<k_{m}=p}\;
\sum_{g_r \in {\mathfrak{F}}_{r}\atop r=1,...,m} 1\le  K^n \ .
\ee
Indeed
\be
\sum_{g_r \in {\mathfrak{F}}_{r}\atop r=1,...,m} 1 = \sum_{r=1}^m \vert {\mathfrak{F}}_{k_r}) \vert 
< 2 (n +2)
\ee
and 
$\sum_{0<k_{1}<...<k_{m}=p} 1 $
is bounded by the number of subsets of $\{0,...,p-1 \}$, hence is bounded by $ 2^p \le 2^n$.
This ends the proof. \qed

\medskip

This allows us to express the self-energy as~:

\begin{multline}
\label{bigformula}
\Sigma (y,z) = \sum_{n=0}^\infty \frac{\lambda^{n+2}}{n!} \int_{\Lambda^n} d^3x_1 ... d^3x_n 
\sum_{\{ a_v \}, \ \{ b_v \}} \sum_{\text{biped structures} \atop \mathcal{B}} \sum_{\text{external fields} \atop \mathcal{EB}} \sum_{\text{clustering tree} \atop \text{structures} \ \mathcal{C}} \sum_{\text{trees} \mathcal{T} \atop \text{over} \ \mathcal{V}} \sum_{\text{field attributions}  \atop \Omega}  \\
\sum_{\{ \sigma_v^j \}}
\sum_{{m-{\rm arch\ systems} \atop 
\big( (f_1,g_1,...,(f_m,g_m))\big) } \atop
{\rm with} \  m \leq p }
\left( \prod_{\ell \in \mathcal{T}} \int_0^1 dw_\ell \right) \left( \prod_{r = 1}^m \int_0^1 ds_r  \right)
\left( \prod_{\ell \in \mathcal{T}} C_{\sigma(\ell)} (f_\ell,g_\ell)\right) 
\\
\left( \prod_{r=1}^m C(f_r,g_r) (s_1,...,s_{r-1})\right)
 \frac{\partial^m \det_{\text{left}, \mathcal{T}}}{\prod_{r=1}^m \partial C(f_r,g_r)} 
\big( \{ w_\ell\}, \{ s_r\}\big) \ .
\end{multline}

The result of the first expansion is however complicated and it is convenient
to select from the arch system an optimized sub-system, called a minimal 1-PI arch system. 
This defines a map $\phi$ which to any $m$-arch system $\cal{A}$ associates a minimal 1-PI 
${\bar m}$ arch-system  $\cal{M}$. 

To define this map we select as first arch of $\cal{M}$
the unique arch in $\cal{A}$  which starts in $\mathfrak{F}_0$ and ends in $\mathfrak{F}_{q_1}$ with $q_1$ maximal. If $q_1=p$ we are done.
If $q_1 \ne p$, we select as second arch  of $\cal{M}$ the unique one in $\cal{A}$  
 which starts in $\cup_{k = 0}^{q_1} \mathfrak{F}_k$ and ends  in $\mathfrak{F}_{q_2}$ with $q_2$ maximal, and so on. 

Starting from the tree $\mathcal{T}$, we have now a minimal arch system 
of lines which completes it into a 1-PI graph. For simplicity, let us first describe these graphs when the arch system $\cal{M}$ has no "coinciding packet" (i.e. no $\mathfrak{F}_k$ contains more than one arch extremity). We have~:

\begin{itemize}
\item the path $P(y,z,\mathcal{T})$,

\item the ${\bar m}$ arches $(f_1,g_1), \ ... \ , \ (f_{\bar m},g_{\bar m})$ of 
$\cal{M}$ completed by the unique path joining $f_r$ to $x_{{k'}_r}$ 
and the unique path joining $g_r$ to $x_{k_r}$ through $\mathcal{T}$,

\item the remaining links which form subtrees of $\mathcal{T}$.

\end{itemize}

\begin{figure}[H]
\centerline{\resizebox{8cm}{!}{\input{3types.pstex_t}}}
\caption{The three kinds of links after a first arch expansion}
\label{threetypes}
\end{figure}
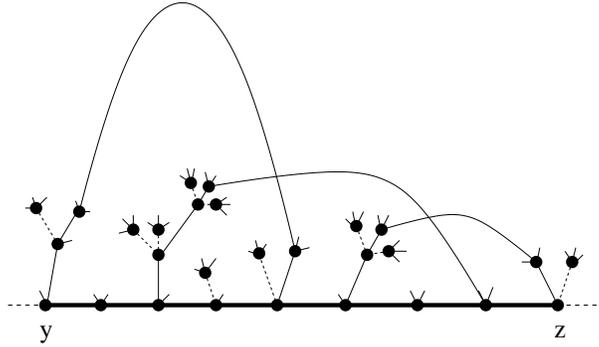

These three kinds of links are illustrated on Figure \ref{threetypes}, where the links of $P(y,z,\mathcal{T})$ are drawn in bold lines, those of the completed arches in "normal" lines and the remaining ones in dashed lines.

When the packet indices are all different, the structure 
of the minimal 1PI-arch-system is therefore the one represented on Figure \ref{archss}~:
\begin{figure}[H]
\centerline{\psfig{figure=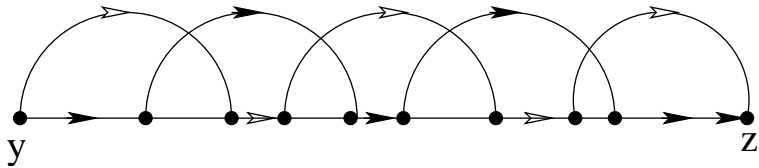,width=10cm}}
\caption{The minimal 1-PI structure without the remaining links of $\mathcal{T}$}
\label{archss}
\end{figure}
\noindent which can also be represented as a kind of "fish", whose borders are shown with corresponding arrows
in the previous figure~:

\begin{figure}[H]
\centerline{\psfig{figure=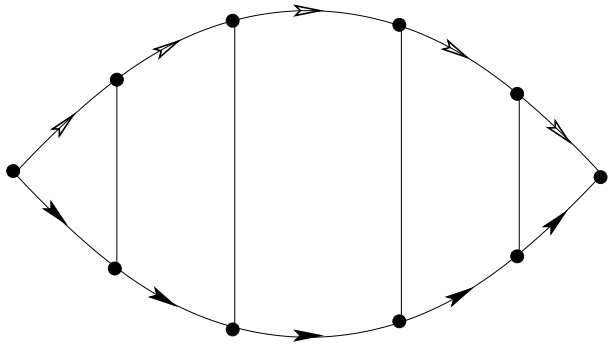,width=8cm}}
\caption{The "fish" structure}
\label{fish}
\end{figure}

Now let us examine the case where the minimal 1PI-arch-system has "coinciding packets", i.e. where the end of some arch and the origin of the next one belong to the same $\mathfrak{F}_k$. We shall distinguish various cases, according to the way the two arches are branching to $\mathfrak{F}_k$. From the vertex $x_k$, apart from the two links of $\mathcal{P} (y,z,\mathcal{T})$, hook two half-lines or lines which are potentially the beginning of two subtrees of $\mathcal{T}$. First, consider the case where $x_k$ has two half-lines. Then the branching of the arches is like on Figure \ref{2halflines} (case 1)~:

\begin{figure}[H]
\centerline{\psfig{figure=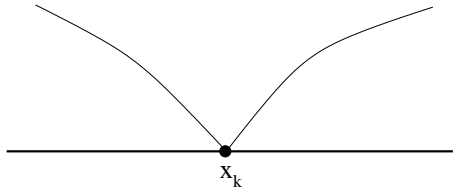,width=6cm}}
\caption{The branching of two arches when $x_k$ has two half-lines (case 1)}
\label{2halflines}
\end{figure}

\begin{figure}[H]
\centerline{\psfig{figure=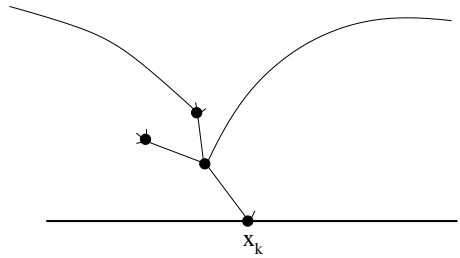,width=6cm}}
\caption{Two arches branching on the same subtree of $\mathcal{T}$ (case 2)}
\label{1halfline2}
\end{figure}

\begin{figure}[H]
\centerline{\psfig{figure=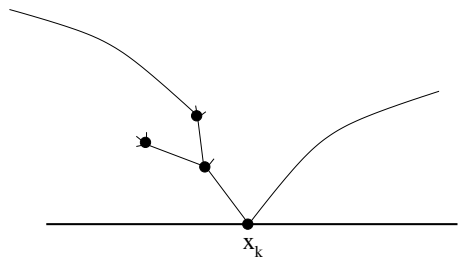,width=6cm}}
\caption{One arch branching on a subtree and the other one on the half-line of $x_k$ (case 3)}
\label{1halfline1}
\end{figure}

If $x_k$ bears a half-line and a subtree of $\mathcal{T}$, we must distinguish two sub-cases~: both arches can hook to the subtree (cases 2), or only one of them can hook to the subtree whereas the other one hooks to the half-line (cases 3). These two situations are pictured on Figures \ref{1halfline2} and \ref{1halfline1}.

\begin{figure}[H]
\centerline{\psfig{figure=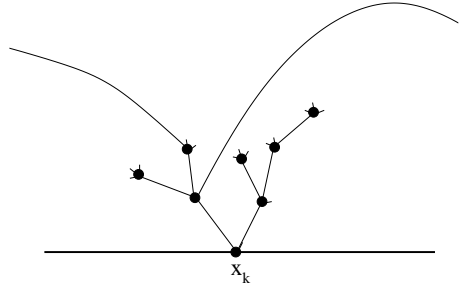,width=6cm}}
\caption{Two arches branching on the same subtree, the other subtree being not touched (case 4)}
\label{2subtrees2}
\end{figure}

\begin{figure}[H]
\centerline{\psfig{figure=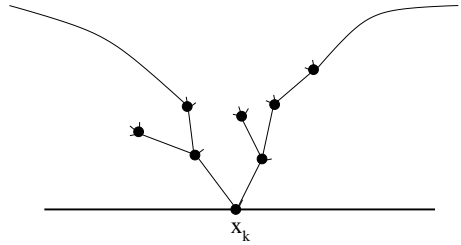,width=6cm}}
\caption{Two arches branching on different subtrees (case 5)}
\label{2subtrees1}
\end{figure}

At last, if $x_k$ is the root of two subtrees of $\mathcal{T}$, we have two sub-cases~: both arches can hook to the same subtree (case 4), or each of them can hook to distinct subtrees (case 5). These two sub-cases are represented on Figures \ref{2subtrees2} and \ref{2subtrees1}.

The inspection of these five cases reveals that the "fish structure"  of Figure \ref{fish} iterates.
Cases 1, 3 and 5 induce a pinch leading to a "new fish"
separated from the previous one by a vertex of reducibility (1-VR). Cases 2 and 4 do not induce any pinch but simply enlarge the "fish". 

\begin{figure}[H]
\centerline{\psfig{figure=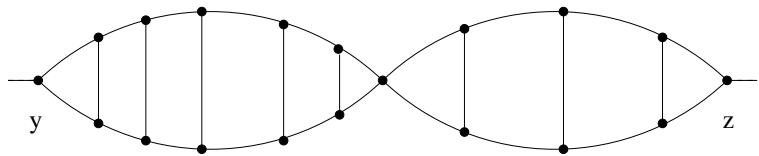,width=10cm}}
\caption{The general 1-PI structure}
\label{doublefish}
\end{figure}

In the end we obtain a sequence of "fishes" separated by vertices of reducibility, as in Figure \ref{doublefish}.

This object is called a fish structure, and it is made of an upper and a lower path, together
with middle bars and middle 1-VR vertices. Any vertex on the upper or lower path
which is neither a middle 1-VR vertex nor on a middle bar is called a ladder vertex.
In the next section we use the minimal 1PI-arch-system as a guide for a 2-PI expansion, 
just like the initial tree ${\cal T}$ was the guide for the 1-PI expansion.

\subsection{2-particle-irreducible arch expansion}

The self-energy $\Sigma_2 (y,z)$ is defined as the sum of the 1-PI contributions, but it has automatically a stronger property in our model~: it is 2-PI and one-vertex irreducible (1-VI). This is just a consequence of the fact that all vertices in our theory have coordination 4. To take advantage of this fact, we devise an additional arch expansion, which derives explicitly more lines out of the determinant. These additional lines, which insure 2-PI, are necessary for the proof of theorem \ref{theortwopoint}. Nevertheless, we must be careful in performing this second arch expansion to respect again the positivity
property so that Gram's bound is not deteriorated, and also to check the analog
of Lemma \ref{lemmeconstr}, that is the constructive character of the expansion.

A naive approach could consist in keeping the definition of the previous "field packets" $\mathfrak{F}_k$ (in which, of course, the fields used in the first expansion are deleted), but this would not select exactly the 2-PI contributions. 

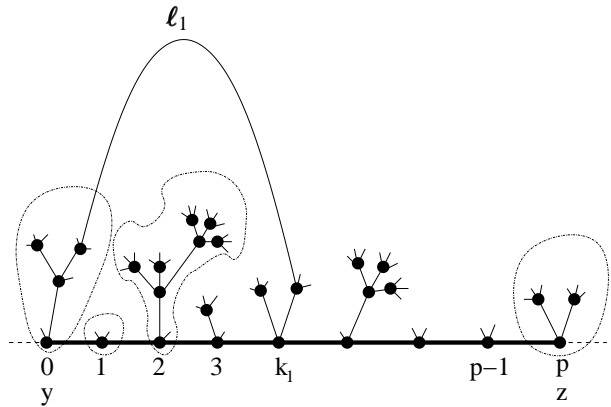
\begin{figure}[H]
\centerline{\resizebox{8cm}{!}{\input{arbre6.pstex_t}}}
\caption{An arch not hooked directly to $y$}
\label{bad_2PI}
\end{figure}

For example, if the first arch of the first expansion is of the type of Figure \ref{bad_2PI}, 
\noindent that is, if the starting field is not hooked directly to the vertex $y$, the second arch expansion could arise as in Figure \ref{bad_2PI2}~:
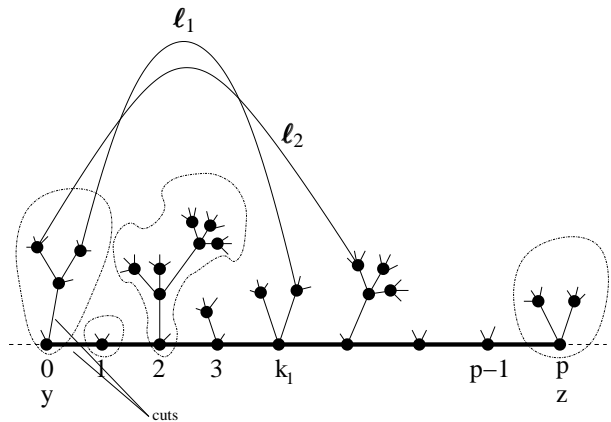
\begin{figure}[H]
\centerline{\resizebox{8cm}{!}{\input{arbre7.pstex_t}}}
\caption{The beginning of a wrong 2-PI arch expansion}
\label{bad_2PI2}
\end{figure}
\noindent and the two cuts indicated on the picture would still disconnect $y$ and $z$. In order to avoid this difficulty, we need to use the general structure of the graph obtained  after the first arch expansion.
to define the new "field packets" and there is a small additional difficulty, which is that
these packets are not totally ordered but only partially ordered in a natural way.

\begin{figure}[H]
\centerline{\psfig{figure=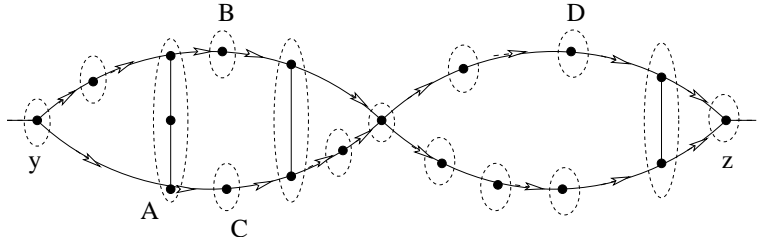,width=10cm}}
\caption{The partial ordering in a multi-fish structure}
\label{doublefisharrow}
\end{figure}

The new definition of the field packets is the following~: a field packet contains either 
all the fields whose path to $y$ first meets the fish structure in a given middle bar, 
or all the fields whose path to $y$ meets the  fish structure at a given
ladder vertex. In the first case we say that we have a "bar packet", in the last case we have a 
"ladder packet". Finally we could also add packets for each middle reducibility vertex, also called
bar packets; although they do not contain any field, it is convenient to introduce them for 
consistency of the partial ordering defined below. 
These packets are shown in Figure \ref{doublefisharrow} as dotted ellipses~: 
in this figure there are 6 "bar packets" and 9 "ladder packets".

Furthermore we have an ordering on these packets, but it is only a partial ordering, noted $\prec$.
If we put arrows from $y$ to $z$ on the two
outer paths in the fishes, packets $A$ and $B$ satisfy $A \prec B$ if and only if one can go from
$A$ to $B$ by a path which does not run against any arrow. 

For instance in Figure \ref{doublefisharrow}, we have $A \prec B \prec D $ and $A \prec C \prec D $
but there is no relation between $B$ and $C$.

To grasp this partial ordering better, we can label the bar packets as $\mathfrak{G}_0,
\mathfrak{G}_1,..., \mathfrak{G}_q $
and label the ladder packets between bar packets $r$ and $r+1$ as
$\mathfrak{G}_{r,a},\; \mathfrak{G}_{r,b},\; ...$ on the upper path and  $\mathfrak{G}_{r,a'},\; \mathfrak{G}_{r,b'},\; ...$ on the lower path.
This is illustrated on Figure \ref{numbering}~:

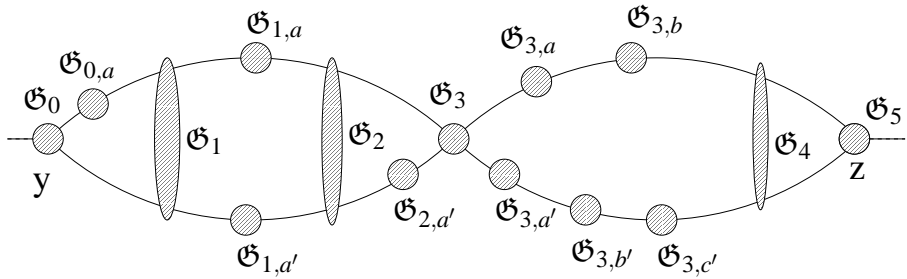
\begin{figure}[H]
\centerline{\resizebox{12cm}{!}{\input{numbering.pstex_t}}}
\caption{The numbering of the field packets for the 2-PI arch expansion}
\label{numbering}
\end{figure}

Once this is done, the 2-PI expansion is carried out in a similar way than before, 
but with a few modifications. We introduce successive interpolation parameters
$s'_1, s'_2, ...$. The first one tests the packet $\mathfrak{G}_0$ with the complement, that is the set
of all later packets in the partial ordering.
Hence the first Taylor expansion step creates a first arch joining this packet $\mathfrak{G}_0$ 
to a bar packet  $\mathfrak{G}_{r}$ or to a ladder packet $\mathfrak{G}_{r, i}$ or $\mathfrak{G}_{r, i'}$
called the first arrival packet. 
Such an arch insures 2-PI only for the block of all packets which are smaller or equal than the arrival packet {\it in the sense of the partial ordering $\prec$}. 

So at second stage we have to launch the second arch from this 2-PI block to the set of 
all the remaining packets not in this block, and so on. 

At any given stage of the induction, the 2-PI block is a "fish-commencing section",
that is either the set of packets smaller or equal to a single given  packet
(of any type $r$, $(r, i)$ or $(r, i')$),  or
the set of packets smaller or equal to one among two ladder packets $(r, i) $ and $  (r, i') $
with same index $r$, one on the lower and the other on the upper part of the fish.

From this block the next arch is launched  to the remaining packets. 
This defines uniquely inductively our expansion.

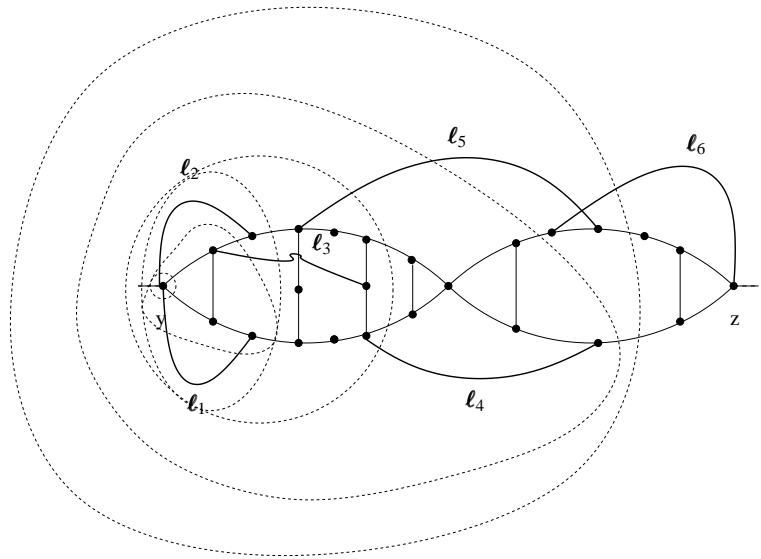
\begin{figure}[H]
\centerline{\resizebox{10cm}{!}{\input{twopipoisson.pstex_t}}}
\caption{A possible arch system for the 2-PI arch expansion}
\label{twopiarches}
\end{figure}
Now the result is a system of arches which insures 2-PI from $y$ to $z$.
On Figure \ref{twopiarches}
we have shown a possible example of such a system.
The arches are represented by bold lines ${\pmb \ell}_1$, ${\pmb \ell}_2$, ... and the corresponding
successive 2-PI blocks are shown by the successive larger and larger dotted
surrounding contours. 

The final result is therefore
given by the same kind of formula than \ref{bigformula}. If we call
the second arch system an $m'$-arch system, we have:
\begin{multline}
\Sigma (y,z) = \sum_{n=0}^\infty \frac{\lambda^{n+2}}{n!} \int_{\Lambda^n} d^3x_1 ... d^3x_n 
\sum_{\{ a_v \}, \ \{ b_v \}} \sum_{\text{biped structures} \atop \mathcal{B}} \sum_{\text{external fields} \atop \mathcal{EB}} \sum_{\text{clustering tree} \atop \text{structures} \ \mathcal{C}} \sum_{\text{trees} \mathcal{T} \atop \text{over} \ \mathcal{V}} \sum_{\text{field attributions} \atop \Omega} \sum_{\{ \sigma_v^j \}}\\
\sum_{ m-{\rm arch\ systems} \atop \bigl( (f_1,g_1), ... (f_m,g_m) \bigr)}
\sum_{m'-{\rm arch\ systems} \atop \bigl( (f'_1,g'_1), ... (f'_{m'},g'_{m'}) \bigr)}  
\left( \prod_{\ell \in \mathcal{T}} \int_0^1 dw_\ell \right) \left( \prod_{\ell \in \mathcal{T}} C_{\sigma(\ell)} (f_\ell,g_\ell)\right) \\
\left( \prod_{r=1}^m \int_0^1 ds_r \right) \ \left( \prod_{r'=1}^{m' }\int_0^1 ds'_{r'} \right)
\left( \prod_{r=1}^m C(f_r,g_r) (s_1,...,s_{r-1})\right) \left( \prod_{r' = 1}^{m'} C({f'}_{r'},{g'}_{r'}) (s'_1,...,s'_{r'-1})\right) \\
 \frac{\partial^{m+m'} \det_{\text{left}, \mathcal{T}}}{\prod_{r=1}^m \partial C(f_r,g_r)\prod_{r'=1}^{m'} \partial C(f'_{r'},g'_{r'}) }\big( \{ w_\ell\}, \{ s_r\} ,  \{ s'_{r'} \}  \big)
\end{multline}
In such a formula the nested sums are other all compatible possibilities, in particular the
$m'$-arch system has to be one of the possible ones that can arise using the fish structure
of the $m$-arch system as the guide for the second expansion. 

This formula
displays explicit 2-PI. Using the fact that vertices have coordination four, 
it also displays explicit 1-VI.
We have to check that it also respects positivity and 
remains constructive, i.e. satisfies an analog of Lemma \ref{lemmeconstr}.

The expansion respects again positivity of the interpolated propagator at any stage,
for the same reasons than the first one, namely all the $s'_{r'}$ interpolations are always performed between a subset of packets and 
its complement, so the final covariance as function of the $s'_{r'}$ parameters
is a convex combination with positive coefficients of block-diagonal covariances. This ensures
that the presence of these $s'_{r'}$ parameters again does not alter Gram's bound
on the remaining determinant.

We need finally to check that the expansion is still constructive. 
Arches system such as those of Figure \ref{twopiarches} obey some constraints.
For two arches ${\pmb \ell}_i$ and ${\pmb \ell}_j$ with $i<j$, the arrival packets $A_i$
and $A_j$ cannot coincide and it is not possible to have
$A_j \prec A_i \;$, hence arrivals respect the {\it partial }
ordering $\prec$. Furthermore let us say that  the arch is of upper type if
the arrival packet is a bar packet with index $r$ or an upper ladder packet $(r, i)$ 
and is of lower type if the arrival packet is a lower ladder packet with index $(r, i')$. Then the
set of arrival points for upper type arches is strictly ordered under $\prec$, and so is the
set of arrival points for lower type arches.

Hence we can fix separately the set of arrival fields, the set of departure fields,
and for each arch for $r' = 1, ..., m'$ whether it is an upper or lower arch.
This choice costs at most $4^{2(n+2)} 2^{m'} \le 2^{5(n+2)}$, 
since the total number of fields is at most $2(n+2)$
(this is not an optimal bound!).
Once this choice is fixed we know exactly the arrival points $g'_{r'}$ for each arch. 
Then the choice of the corresponding departure points is determined using the $s'_{r'}$
parameters exactly as in Lemma \ref{lemmeconstr}, where the ${\mathfrak{E}}_r$ are now the sets of
strictly growing "commencing sections", that is the successive regions surrounded by dotted contours
in Figure \ref{twopiarches}, and the numbers $e_i = \vert {\mathfrak{E}}_i - {\mathfrak{E}}_{i-1} \vert $  
are now the total number of fields hooked to the region
between two successsive contours with labels $i-1$ and $i$. 
Therefore Lemma \ref{lemmeconstr} also holds for
the $m'$ arch system.

\subsection{Three disjoint paths}

By the previous double arch expansion, we have explicitly displayed the 2-PI structures contributing to the self-energy. The advantage of this expansion is that we have now at our disposal more explicitly derived
links, which can be used to bound in a better way the integrations $\int \prod_{v} dx_{v,0}dx_{v,+}dx_{v,-}$. Integrating the vertices positions in the standard way using the
decay of the lines of a single tree connecting all the vertices is (apparently) not sufficient to obtain the requested bounds of theorem V.1. Thanks to the 2-PI structure extracted by the double arch expansion, we are going to forge a better scheme of integration.

We need a theorem (in fact, two versions of the same theorem) known as Menger's theorem. Roughly speaking, it states that in a $p$-particle-irreducible graph, there exists (at least) $p+1$ line-disjoint paths joining two given vertices. A cautious statement of this result is the following one~:

\medskip
\noindent \textbf{Theorem ("edge version" of Menger's theorem)~:}

\textit{Let $G$ be a graph, $u$ and $v$ two distinct vertices of $G$. Suppose that $u$ and $v$ cannot be disconnected by the deletion of $p$ lines (edges) of $G$, for $p \in \mathbb{N}$. Then there exists $p+1$ line-disjoint paths joining $u$ and $v$ through $G$.} \qed

\medskip

Two (or more) paths $P_1$ and $P_2$ are said line-disjoint if $P_1 \cap P_2 = \emptyset$ (remember that a path is by definition a set of lines). We stress the fact that these paths whose existence is insured by the edge version of Menger's theorem may go across some identical vertices; in other words, they are not necessarily \textit{vertex-disjoint}, even if we take away the end vertices $u$ and $v$.

But there exists another version of Menger's theorem~:

\medskip
\noindent \textbf{Theorem ("vertex version" of Menger's theorem)~:}

\textit{Let $G$ be a graph, $u$ and $v$ two distinct vertices of $G$. Suppose that $u$ and $v$ cannot be disconnected by the deletion of $p$ vertices ($p \in \mathbb{N}$). Then there exist $p+1$ internally vertex-disjoint paths joining $u$ and $v$.} \qed

\medskip

We say that two paths $P_1$ and $P_2$ are internally \textit{vertex-disjoint} if $P_1$ and $P_2$, once deprived from their end vertices, have no vertex in common. For more details about these two versions of Menger's theorem, the reader may consult \cite{Bondy} or any textbook on graph theory. Although
Menger's theorems are very simple, their proof is quite subtle. They can be seen as corollaries of a famous powerful theorem of graph optimization, the so-called "max flow-min cut theorem" \cite{Bondy}.

It is easy to give examples of 2-PI graphs for which the previous theorem naturally holds, but in which it is impossible to exhibit three vertex-disjoint paths, for instance the graph of Figure \ref{nicegraph}~:

\begin{figure}[H]
\centerline{\psfig{figure=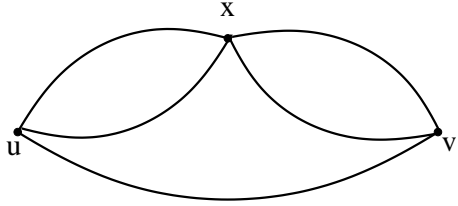,width=6cm}}
\caption{A 2-PI graph with no triplet of vertex-disjoint paths joining $u$ and $v$}
\label{nicegraph}
\end{figure}

Note also that the theorem does not state that the set of the paths is unique in general. Unicity can be insured only in the very special case of graphs having the (rather trivial) structure of Figure \ref{structri}~:

\begin{figure}[H]
\centerline{\psfig{figure=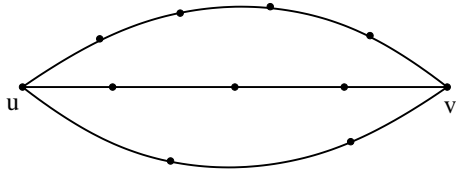,width=6cm}}
\caption{A graph with a unique triplet of line-disjoint paths joining $u$ and $v$}
\label{structri}
\end{figure}

\begin{figure}[H]
\centerline{\psfig{figure=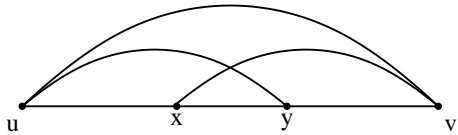,width=6cm}}
\caption{A 2-PI graph for which Menger's theorem is not trivial}
\label{gluop}
\end{figure}

But if the graph $G -\{u,v\}$ has vertices linked to more than two neighbors it is possible to find several sets of three line-disjoint paths connecting $u$ and $v$. Finally, we remark that these paths cannot be determined naively and independently of the other ones. For example, in the graph of Figure \ref{gluop}~,
\noindent if we choose the first two paths as being $\{\{u,v\}\}$ and $\{\{u,x\},\{x,y\},\{y,v\}\}$, we cannot find a third one. Thus the result of the theorem is quite subtle and not totally obvious.

The set of lines we derived explicitly thanks to our initial tree expansion and our two successive arch expansions is by construction 2-PI in the channel $y - z$. Then a straightforward application of the edge version of Menger's theorem insures that, if we call $G$ the graph whose vertices are $\mathcal{V}$ and lines those of $\mathcal{T}$ plus the ones explicitly derived by the two arch expansions, there exist (at least) 3 line-disjoint paths $P_1$, $P_2$  and  $P_3$ joining $y$ to $z$. 

From now on for vertex integration purposes we use only the
the lines in $L= \mathcal{T} \cup P_1\cup P_2\cup P_3$, hence forget any arch line not
in $P_1\cup P_2\cup P_3$  and the  remaining fields in the determinant or remaining lines.
Remark that the union $ \mathcal{T} \cup P_1\cup P_2\cup P_3$ is not necessarily disjoint, since some lines of the $P_i$'s may belong to $\mathcal{T}$.

\section{Ring Construction}

In this section the lines scales enters the picture. Out of the lines of $L$ we shall extract
a subset, called a ring, which is the union of {\it two} line-and-vertex-disjoint paths from 
$y$ to $z$. This ring has to satisfy Lemma \ref{lemmacrux} below and its extraction depends
therefore on the Gallavotti-Nicol\`o tree structure associated to the
scales assignments over all lines and fields.

We consider therefore the forest $\cal F$ 
of those connected parts or nodes $\Gamma$ of the Gallavotti-Nicol\`o tree associated
to the scale decomposition $i$ (including the initial bare vertices, with four legs).
The full two point contribution $G$ that we analyze is itself such a node $\Gamma_0 =G$. 
Recall that for any such $\Gamma$ and any pair of its external legs, 
there exists a unique path (eventually empty!) in $\mathcal{T} \cap \Gamma$ joining 
the two vertices two which these two external lines are hooked. We call
this path the "tree shortcut" for the pair. This is because $\mathcal{T}\cap \Gamma$  
is a tree of $\Gamma$. Since we are studying
primitively divergent two point subgraphs, any $\Gamma$ 
except $G$ itself has at least 4 external legs. (If $\Gamma $ contains $y$ or $z$, we count the corresponding external lines of $G$ as external legs of  $\Gamma $).

\begin{lemma}
There exists a ring $R \subset L$ which is the union of two line-and-vertex-disjoint 
paths from $y$ to $z$, with the additional property 
that for any $\Gamma \in {\cal F}$,  at least 2 external legs of  $\Gamma $ are not in the ring $R$.
\label{lemmacrux}
\end{lemma}
{\bf Proof}

An element $\Gamma $ is called a "cut" if removing it separates $y$ from $z$, or in other words if every path in $L$ from $y$ to $z$ touches $\Gamma$. 
It is called "contractible" if it is not a cut. 

We consider the set $S$ of all maximal  contractible elements in $\cal F$ (by our convention
they can be ordinary bare vertices). Elements of $S$ must be all disjoint by the forest character
of $\cal F$.

We reduce each element of $S$ to a point, that is we ignore
the interior of any element of $S$, 
and keep all the elements of $S$ plus all the lines and determinant fields attached to them
connected as before. In this way we obtain a new graph $G'$, which has generalized vertices 
with 4 legs or more, in particular it has one such vertex for each element of $S$. It must still have three line-disjoint 
paths $P'_1$, $P'_2$ and $P'_3$, made of those lines in $P_1$, $P_2$ and $P_3$ which were not internal to any contractible element of $\cal F$. The graph $G'$ is 
therefore still 2-lines irreducible in the channel $y \to z$.

\begin{figure}[H]
\centerline{\psfig{figure=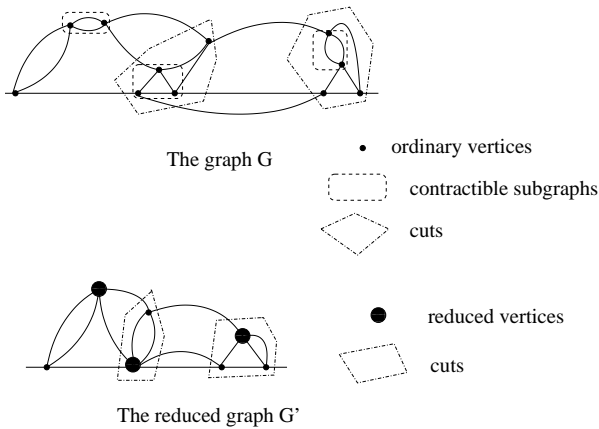,width=8cm}}
\caption{The process of contraction}
\end{figure}

$G'$ is also one-vertex irreducible in the 
chanel $y \to z$, since by definition for any vertex $v$ of $G'$ distinct from $y$ and $z$
there is a path in $L$ from $y$ to $z$ which avoids  
$v$, so the corresponding reduced path in $G'$ also avoids the vertex $v$.

By the vertex-version of Menger's theorem, 
there is therefore a {\it ring} $R'\subset G'$ in this graph, namely a subset of lines which is
the union of two $vertex$ disjoint paths $R'_1$ and $R'_2$ from $y$ to $z$. 

We consider now the graph $G'-R'$. It must connect $y$ to $z$. Otherwise 
there would be a connected component $C(y)$ of $G'-R'$ containing $y$ and not $z$, and
removing the two last exits of $R'_1$ and $R'_2$ from that component would disconnect
$G'$, hence $G'$ would not be 2-PI in the channel $y\to z$.

Therefore there exists a path $R'_3$ from $y$ to $z$ in $G'$ entirely line-disjoint from the ring
$R'$.

\begin{figure}[H]
\centerline{\psfig{figure=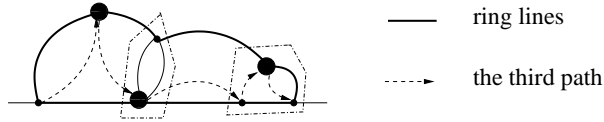,width=8cm}}
\caption{The ring in $G'$}
\end{figure}

We complete the ring $R'$ of $G'$ into a true ring $R$ of $G$ by 
adding, for every non-bare vertex of 
$G'$ touched by the ring the shortcut between the entrance and exit
in $\mathcal{T}$. Clearly this defines $R\subset L$ in a unique way.

Let us check that $R$ has the desired property. It is obvious for 
$\Gamma$'s which are contractible. Indeed 
\begin{itemize}
\item either they are 
maximal contractible, in which case they are touched by only one 
of the two vertex-disjoint paths $R'_1$ and $R'_2$ of the ring, and the number
of their external legs in the ring $R'$ is at most two, the entrance and exit of that path.
We are done, since $\Gamma$ has at least four external legs, of which only two belong to $R$
(the ones in $R-R'$ are internal to $\Gamma$ or disjoint from $\Gamma$).

\item or  $\Gamma$ is not maximal contractible, hence strictly inside 
a reduced vertex of $G'$. Then recall that the ring $R$ is made of a corresponding tree shortcut in 
$\mathcal{T}$. Again it can touch $\Gamma$ only twice (it cannot enter and reexit, since $\mathcal{T}$, w has no loops; this is due to the key
property of $\mathcal{T}$, whose restriction to any $\Gamma$ node in the GN tree is a spanning tree in $\Gamma$). We conclude in the same way.
\end{itemize}

Therefore we have to consider $\Gamma$'s which are cuts.
But such $\Gamma$ appear as subgraphs $\Gamma'$ in $G'$
which must be still cuts of $G'$. Therefore they must be touched by $R'_3$.
Following $R'_3$, its first entrance into $\Gamma'$ (when $y \not \in \Gamma$) and its
last exit out of $\Gamma'$ (when $z \not \in \Gamma$) give external legs of $\Gamma'$
which do not belong to $R'$, hence two external legs of $\Gamma$
which do not belong to $R$. \qed

\subsection{Ring Sector}

Let us now return to the bound on the primitively divergent self energy contribution with
cutoff $r_{max}$, namely $ |\Si ^{\le r_{max}}_{2,pr} (y,z)| $. 

There is a first scale $r_\mathcal{T}$ at which $y$ and $z$ fall into a common connected component
of the GN tree. It is the largest index on the initial path in $\mathcal{T}$ from $y$ to $z$. Let us call $r_R$ the first $r$ index at which the ring connects $y$ and $z$.
Obviously since $\mathcal{T}$ is optimized with respect to the $r$ indices, we have 
$r_\mathcal{T} \le r_R$. $r_R$ can be expressed as a minimax 
over the two disjoint paths $P_{R,1}$ and $P_{R,2}$ 
which compose the ring:
\be
r_{R} = \min_{j=1,2} r_{R,j} \ ; \   r_{R,j} =  \ max_{k \in P_{R,j}} r(k) 
\ee
Obviously we have $r_{R} \le r_{max}$. 

In the same vein we should define a (generalized) sector $\bar \si_R$ associated to the ring $R$ and the tree $\mathcal{T}$.
It is a triplet $(i_{R,\mathcal{T}}, s_{+,R}, s_{-,R})$ depending on the sector attributions of the lines of the tree $\mathcal{T}$ and
of the ring $R$ we have just built. $s_{+,R}$ and $s_{-,R}$ are also minimax 
of the corresponding indices over the two disjoint paths $P_{R,1}$ and $P_{R,2}$ 
which compose the ring. More precisely
\be
s_{+,R} = \min_{j=1,2} s_{+, R,j} \ ; \   s_{+, R,j} = \max_{k \in P_{R,j}}    s_+(k) 
\ee
\be
s_{-,R} = \min_{j=1,2} s_{-, R,j} \ ; \   s_{-, R,j}  \max_{k \in P_{R,j}} s_-(k) . 
\ee
The index $i_{R,\mathcal{T}}$ is optimized both over $\mathcal{T}$ and $R$. More precisely we define,
if $P(y,z,\mathcal{T})$ is the unique path from $y$ to $z$ in $\mathcal{T}$:
\be
i_{\mathcal{T}} = \max_{k \in P(y,z,\mathcal{T})} i(k) 
\ee
\be
i_{R} = \min_{j=1,2}   i_{R,j}  \ ; \    i_{R,j} = \max_{k \in P_{R,j}} i(k) 
\ee
\be
i_{R,\mathcal{T}} = \min  \{ i_{R} , i_{\mathcal{T}} \}
\ee

Using the relations $0 \le s_{\pm} \le i $ for ordinary sectors, one has 
$s_{\pm, R, j} \le i_{R,j}  \le r_{R,j}$, hence  $0 \le s_{\pm,R} \le r_{max}$. Furthermore $i_\mathcal{T} \le r_\mathcal{T} \le r_{max}$
so that the three indices $i_{R,\mathcal{T}}$,   $s_{+,R}$ and $s_{-,R}$ being all bounded
by $r_{max}$ are indeed those of a generalized 
sector $\bar \si_{R,\mathcal{T}}$ of $\Si _{r_{max}}$, in the sense of Section \ref{maintheo}. 
We define the associated $r$ index of this generalized sector as $r_{R,\mathcal{T}}$:
\be r_{R,\mathcal{T}} \equiv \frac{i_{R,\mathcal{T}} + s_{+,R} + s_{-,R}}{2}
\ee

We also define the scaled distance for that ring sector $\bar \si_{R,\mathcal{T}} 
= (i_{R,\mathcal{T}}, s_{+,R}, s_{-,R})$
as 
\be 
d_{R,\mathcal{T}} (y,z) =d_{i_{R,\mathcal{T}},s_{+,R}, s_{-,R}}(y,z)  
\ee

\section{Power counting}

Everything is now prepared for the bounds. We do not repeat all details but concentrate on what is
new with respect to \cite{R1}).

We introduce all the momentum constraints $\chi_j (\si)$ for all the vertices
of the primitively divergent self energy contribution. After that 
we apply Gram's bound on the remaining determinant. This replaces the remaining determinant by a product over its entries of the corresponding power counting factor (see \cite{R1}).

We shall first  perform the spatial integration over the positions of internal vertices,
using the propagators decay and the fields and propagators prefactors. This is really
power counting. Then we shall perform the sector sums, using the coupling constants,
which is a kind of logarithmic power counting.

The spatial integration are themselves divided in two steps. We write:

\be  |\Si ^{\le r_{max}}_{2,pr} (y,z)| \le \sum _{n}\frac{ (c\la )^n}{n!} \sum_{\cal T, R... } \sum_{\si } 
\prod_j \ch_j (\si) I_{1,n}  (y,z) I_{2,n} (y,z,x_{j,\pm} )
\ee
\be |y_+ - z_+|.|y_- - z_-|
|\Si ^{\le r_{max}}_{2,pr} (y,z)| \le \sum _{n}\frac{ (c\la )^n}{n!} \sum_{\cal T, R... } \sum_{\si } 
\prod_j \ch_j (\si) I_{1,n,\pm}  (y,z) I_{2,n} (y,z,x_{j,\pm} )
\ee
\be |y_0 - z_0|.
|\Si ^{\le r_{max}}_{2,pr} (y,z)| \le \sum _{n}\frac{ (c\la )^n}{n!} \sum_{\cal T, R... } \sum_{\si } 
\prod_j \ch_j (\si) I_{1,n,0}  (y,z) I_{2,n} (y,z,x_{j,\pm} )
\ee

In $I_{2,n} (y,z,x_{j,\pm} )$ we keep 
the positions of $y$, $z$ and the {\it spatial} positions of the {\it ring} vertices
$x_{j,\pm } $ {\it fixed} and integrate other all the remaining positions. 
To pay for all these integrations we put in $I_{2,n} (y,z,x_{j,\pm} )$ 
a fraction (say 1/2) of the decay of every line in $L$, all the determinant fields prefactors
and the line prefactors for the line not in the ring. Hence:
\be
I_{2,n}(y,z, x_{j,\pm}) =   \int \prod_{j=1}^{p} dx_{0} \prod_{v \not\in R} d^3x  
\prod_{f \not\in R} M^{-r_f/2 -l_f /4}  \prod_{k \in L } e^{-c.d_{\si(k)}^{\alpha }/2}
\ee
Then in $I_{1,n}  (y,z)$ we gather the remaining factors and integrations.
We first prove a uniform bound on   $I_{2,n} (y,z,x_{j,\pm} )$ independent of the fixed
positions $y,z, x_{j,\pm}$:

\begin{lemma}
The following bound holds:
\be I_{2,n}(y,z,x_{j,\pm} )   
\le K^n  M^{-r_\mathcal{T}}  \prod_{f \not\in R} M^{ -l_f /4} 
\prod_{v \not \in R \atop l, l'\ {\rm  hooked\  to\ } v} M^{- \sum | r_v^l - r_v^{l'} |/18 }
\label{boundnotring}
\ee
\end{lemma}

\noindent{\bf Proof} 

At fixed $+$ and $-$ positions for the ring vertices we integrate
\begin{itemize}
\item all $x_0$ positions with the $\mathcal{T}$ propagators decay

\item all $x_{\pm}$ positions for the vertices not in the ring
with the $\mathcal{T}-R$ propagator decays. 
\end{itemize}
Remark that a vertex $v$ not in the ring integrated with a
tree line of scale $r_v$ costs exactly $M^{2r_v}$, whether when $v$ is in the ring, its
integration over the 0 direction costs only $M^{i_v} = M^{r_v -l_v/2}$ when integrated with a line 
of indices $i_v, l_v$ and $r_v= i_v + l_v/2$.

To compute the result we divide as usual every factor $M^{2r}$, $M^r$ or $M^{-r/2}$
as a product over all scales and we collect everything scale by scale. Following the previous section,
we should distinguish the connected components $\Gamma$ which have empty intersection with the ring, also called {\it ring-disjoint}, and those which contain at least one vertex of the ring,
called  {\it ring-intersecting}. Among these one should also distinguish those who contain
neither $y$ nor $z$, called {\it y and z disjoint}, and those who contain $y$ or $z$ or both. 
There is then at each scale a factor $M^2$ to pay for each {\it ring-disjoint} component (corresponding
to one particular vertex which plays the role of a center of mass for that component, which is integrated
both on time and spatial position); a factor  $M$  to pay for each {\it ring--intersecting, y-z disjoint} component, for which only the time position of a ring vertex has to be integrated, and no factor
to pay for the components containing $y$ or $z$ or both, since $y$ and $z$ are fixed. Therefore
we get

\bqa && I_{2,n}(y,z,x_{j,\pm} )   
\le K^n \prod_{f \not\in R} M^{-r-f/2} \prod _{v\not\in R} M^{2r_v} \prod _{v \in R} M^{r_v-l_v/2}
\le  K^n \prod _{v \in R} M^{-l_v/2}
\prod_{\Gamma\  {\rm ring-disjoint}}  M^{ 2- e(\Gamma)/2} 
\nonumber \\
&&\prod_{\Gamma \ {\rm ring-intersecting}, \ 
y {\rm \ and\ } z\ {\rm disjoint}}
M^{ 1- e'(\Gamma)/2}  \prod_{\Gamma  \ {\rm containing }\ y {\rm \ or\ } z } M^{- e'(\Gamma)/2}
\label{boundfine}
\eqa
where $e' (\Gamma)$ for a connected component $\Gamma$ is the number of external fields
of $\Gamma$ {\it not} in the ring.

Using the previous section we know that for every connected component $\Gamma$ which is
$y\cup z$ disjoint, $e' (\Gamma)\ge 2 $; and since we consider a primitively divergent contribution,
$e (\Gamma)\ge 4 $. Furthermore for every connected component containing $y$ or $z$ but not both, we have $e' (\Gamma)\ge 1$, so that we have a decay factor $M^{-1}$ from the first scale $r=0$ until at least the first scale $r_\mathcal{T}$ at which $y$ and $z$ become connected in $\mathcal{T}$.

Hence following the usual argument as in \cite{R1} we get
\bqa
&& \prod_{\Gamma\  {\rm ring-disjoint}}  M^{ 2- e(\Gamma)/2}  
\prod_{\Gamma \ {\rm ring-intersecting}, \ y {\rm \ and\ }z\ {\rm disjoint}} M^{ 1- e'(\Gamma)/2}  
\prod_{\Gamma  \ {\rm containing }\ y {\rm \ or\ } z } M^{- e'(\Gamma)/2} 
\nonumber \\
&&
\le  M^{-r_\mathcal{T}}\prod_{v \not \in R \atop l, l'\ {\rm  hooked\  to\ } v} M^{- \sum | r_v^l - r_v^{l'}|/18}
\eqa
which completes the proof of the Lemma.
\qed

We treat now the power counting of the ring lines and the space integration of 
the ring vertices together in $I_1$.  We have (recalling that the internal vertices
of the ring have positions $x_1$, .... , $x_p$):

\begin{lemma}
For some constant $K$
\begin{multline}
I_{1,n}(y,z) =  \int \prod_{j=1}^{p} dx_{j,+} dx_{j,-} \prod_{k\in R} M^{- (r + l/2)(k)}\prod_{k\in L} e^{-c . d_{\si(k)}^{\alpha }/2} \\
\le K ^p   M ^{-s_{+, R,1} -s_{-, R,1} -s_{+, R,2} -s_{-, R,2}}
 e^{- c.d_{R,\mathcal{T}}^{\alpha} (y,z)}    \label{firstbou}
\end{multline}
\begin{multline}
I_{1,n,\pm}(y,z) =  \int \prod_{j=1}^{p} dx_{j,+} dx_{j,-}  |y-z|_+ |y-z|_- \prod_{k\in R} M^{- (r + l/2)(k)}
\prod_{k\in L} e^{-c.d_{\si(k)}^{\alpha }/2} \\
\le K^p   M ^{-s_{+, R}  -s_{-, R}}  e^{- c.d_{R,\mathcal{T}}^{\alpha} (y,z)} \label{secondbou}
\end{multline}
\bqa I_{1,n,0}(y,z) &=&  \int \prod_{j=1}^{p} dx_{j,+} dx_{j,-}  |y_0 -z_0| \prod_{k\in R} M^{- (r + l/2)(k)}
\prod_{k\in L} e^{-c.d_{\si(k)}^{\alpha }/2}
\nonumber\\
&\le & K^p   M ^{-s_{+, R,1} -s_{-, R,1} -s_{+, R,2} -s_{-, R,2}}  M^{i_{R,\mathcal{T}}}  \label{thirdbou}
e^{- c.d_{R,\mathcal{T}}^{\alpha} (y,z)}
\eqa
\end{lemma}

\noindent{\bf Proof} 
We use simply the triangular inequality and a fraction the decay of the ring lines or of the $\mathcal{T}$ lines to get the decay $e^{- c.d_{R,\mathcal{T}} (y,z)}$. We keep an other fraction of the
ring lines decay to perform the integration over positions of the ring vertices.
To check the prefactor we remark that
we can separately optimize the $+$ and $-$ integration. The + integration 
consumes all the lines $M^{-s_+}$ prefactors except two, namely the smallest ones on the
two paths of the ring, which are $M^{-s_{+, R,1}}$ and $ M^{-s_{+, R,2}}$. 
Finally in the second bound (\ref{secondbou})
the $|y-z|_+$ factor consumes the largest of these two factors, so we
keep the best factor $M ^{-s_{+, R}}$ in the bound. The same 
is true with the - integrations.
In the third bound (\ref{thirdbou})
we keep both factors $M ^{-s_{\pm, c,1}}$ and  $M ^{-s_{\pm, c,2}}$
but have to pay $M^{i_{R,\mathcal{T}}}$ for the  $|y-z|_0$ factor. 
Here it was important to use the decay of all the lines of $\mathcal{T}$, not only of $R$.
\qed

Let us now remark that in (\ref{boundnotring})we have $M^{-r_\mathcal{T}} \le M^{-i_\mathcal{T}}$.
Let us combine this factor with the other main power counting
factors $ M ^{-s_{+, R,1} -s_{-, R,1} -s_{+, R,2} -s_{-, R,2}} $ of  (\ref{firstbou}),
$ M ^{-s_{+, R}  -s_{-, R}}$ of (\ref{secondbou}) and  
$M ^{-s_{+, R,1} -s_{-, R,1} -s_{+, R,2} -s_{-, R,2} + i_\mathcal{T}} $ of  (\ref{thirdbou}).

For any regular sector $i \le  s_+ + s_- +2 $
we have $i_R \le  \max_{k\in R} s_{+}(k) + \max_{k\in R} s_{-}(k) + 2$, hence
\be
s_{+, R,1} + s_{-, R,1} + s_{+, R,2} + s_{-, R,2} = s_{+, R} + s_{-, R} + \max_{k\in R}
s_{+}(k) + \max_{k\in R} s_{-}(k) 
\ge s_{+, R} + s_{-, R}  + i_{R,\mathcal{T}}  - 2  \label{bibou}
\ee
so that we obtain from (\ref{thirdbou})
\be M ^{-s_{+, R,1} -s_{-, R,1} -s_{+, R,2} -s_{-, R,2}}   M^{-r_{\mathcal{T}} }  \le 
M ^{-s_{+, R} -s_{-, R} -i_{R,\mathcal{T}} + 2} = K M ^{-2r_{R,\mathcal{T}}} 
\ee 
Similarly combining the factors in (\ref{secondbou}) with $M^{-r_{\mathcal{T}} } \le M^{- i_{R,\mathcal{T}}}$ we have 
\be   M ^{-s_{+, R}  -s_{-, R}}  M^{-r_\mathcal{T}} \le M ^{-2r_{R,\mathcal{T}}} 
\ee
and finally we also have from(\ref{bibou})
\be s_{+, R} +s_{-, R} + i_{R,\mathcal{T}} \le 2 ( \max_{k\in R} s_{+}(k) + \max_{k\in R} s_{-}(k) ) +2
\ee
so that, since $i_{R,\mathcal{T}} \le r_\mathcal{T}$
\be   3 (s_{+, R} +s_{-, R} + i_{R,\mathcal{T}}) \le 2 (s_{+, R,1} + s_{-, R,1} +s_{+, R,2} +s_{-, R,2}
+ r_\mathcal{T}) .
\ee
Therefore in (\ref{firstbou}) we have
\be M ^{-s_{+, R,1} -s_{-, R,1} -s_{+, R,2} -s_{-, R,2}} M^{-r_\mathcal{T}} \le M ^{-(3/2)
(s_{+, R} +s_{-, R} + i_{R,\mathcal{T}})} = M ^{-3r_{R,\mathcal{T}}} 
\ee

It remains to
combine the factor $ \prod_{f \not \in R} M^{ -l_f /4} $
in (\ref{boundnotring}) with the factor $ \prod_{v \in R} M^{ -l_v /2} $ in (\ref{boundfine}) 
to reconstruct the factor 
$ \prod_{f } M^{ -l_f /4} $ for all fields and to collect everything to obtain

\bqa
|\Si ^{\le r}_{2,pr} (y,z)| &\le & \sum _{n}
\frac{ (K\la )^n}{n!} \sum_{\cal T, R... } \sum_{\si } 
\prod_j \ch_j (\si)     \prod_{f } M^{ -l_f /4} 
\nonumber\\
&&\prod_{v \not \in R \atop l, l'\ {\rm  hooked\  to\ } v}
M^{- \sum | r_v^l - r_v^{l'} |/18 }    M ^{- 3r_{R,\mathcal{T}}} 
 e^{- c.d_{R,\mathcal{T}}^{\alpha} (y,z)}
\eqa
\bqa
 &&
 |y-z|_+ |y-z|_-|\Si ^{\le r}_{2,pr} (y,z)| \le \sum _{n}\frac{ (c\la )^n}{n!} \sum_{\cal T, R... } \sum_{\si } 
\prod_j \ch_j (\si)     \prod_{f } M^{ -l_f /4}
\nonumber\\
&&
 \prod_{v \not \in R \atop l, l'\ {\rm  hooked\  to\ } v}
M^{- \sum | r_v^l - r_v^{l'} |/18 }    M ^{-2r_{R,\mathcal{T}}}  e^{- c.d_{R,\mathcal{T}}^{\alpha} (y,z)}
\eqa
\begin{multline}
 |y_0-z_0| |\Si ^{\le r}_{2,pr} (y,z)| \\
\le \sum _{n}\frac{ (c\la )^n}{n!} \sum_{\cal T, R... } \sum_{\si } 
\prod_j \ch_j (\si) \prod_{f } M^{ -l_f /4} 
\prod_{v \not \in R \atop l, l'\ {\rm  hooked\  to\ } v}
M^{- \sum | r_v^l - r_v^{l'} |/18 }    M ^{-2r_{R,\mathcal{T}}} 
e^{- c.d_{R,\mathcal{T}}^{\alpha} (y,z)}
\end{multline}

We have now to sum up over all the sector indices. This is a logarithmic power
counting problem. The sum over all sectors assignments for the vertices 
and 4-legged components (called {\it quadrupeds}" 
not in the ring can be performed exactly as in \cite{R1}.
There it was proved that these sums cost at most one factor $\log T$ per vertex and
one factor $\log T$ per quadruped.

We have also now to check that the bounds also hold for 
the sums over the sectors of the ring vertices.
The ring intersecting maximal connected components have now
become apparently logarithmically divergent, even if they are not quadrupeds,
because for them we only know that 
$1- e'(\Gamma)/2 \le 0 $. So fixing their largest internal $r$ scale once their 
first external $r$ scale is known costs one factor $\log T$. 
Once all $r$ scales are fixed, we have to sum over the auxiliary indices $l$ and 
the $s_+$ and $s_-$ indices, subject to the constraint $s_+  + s_- = r +l/2$.
This is again done at a cost of one factor $\log T$ per vertex using the momentum
conservation rule and the auxiliary decay factor $ \prod_{f } M^{ -l_f /4}$ 
as in \cite{R1}, Lemma 5. 

The result is at most a factor $\log T^{2n-1}$ for fixing all sectors, like in \cite{R1}.
This is why we get analyticity only in a domain $\la \le c /\log^2 T$.
 
Finally we have to sum over the tree $\mathcal{T}$, the arch constructions and over $n$.
This is standard or explained above. In this way the proof of Theorem \ref{theortwopoint} is achieved
for {\it primitively divergent} contributions. In the next section we use an induction to extend
this proof to the general case. 

\section{Chains of bipeds}

\begin{figure}[H]
\centerline{\psfig{figure=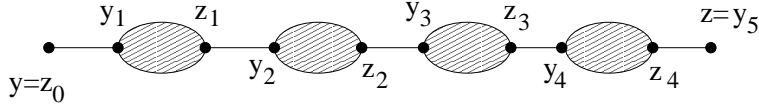,width=10cm}}
\caption{A chain of bipeds}
\label{chainbips}
\end{figure}

Using  Theorem \ref{theortwopoint} we are now in a position to bound
a maximal chain $Chain_r$  of primitively divergent 1PI bipeds 
$B_1$, ...., $B_q$ with fixed ends $y=z_0$ and $z=y_{q+1}$, such as the one of Figure \ref{chainbips},
with the $q+1$ ordinary lines in a sector $\si$ of scale $r$, not yet summed.
The two external vertices of biped  $B_j$ are called $y_j$ and $z_j$.

By momentum conservation, the sectors of all the one particle reducible lines
between the 1PI bipeds must have all the same sector $\sigma$ (or a neighboring one).
We shall compare the bound for such a chain, when all external positions
$y_j,z_j$, $j=1,...,q$ are summed, to that of a regular line.

We have first to evaluate the action of the $\tau^*$ operators  in (\ref{formuleyx})
on the external lines.
We find
\begin{multline}
(\partial_+  + i )(\partial_-  + i )
C_{\si} \Big[\big(x_0, sy_+  + (1-s)x_+, ty_-  + (1-t) y_+\big) , z\Big]  \\
\le M^{-2r(\si) -l(\si)}  \sup\Big[e^{-d_{\si} ^{\al}(x,z)} ,  e^{-d_{\si} ^{\al}(y,z)} \Big]   \label{renprop}
\end{multline}
\be  (\partial_0   - i\pi T ) C  \Big[\big(ty_0+(1-t)x_0, y_+,y_-\big) , z\Big]  
\le M^{-2r(\si)}  
\sup\Big[e^{-d_{\si} ^{\al}(x,z)} ,  e^{-d_{\si} ^{\al}(y,z)} \Big] \label{renproz}
\ee
We should take into account that there are $q+1$ ordinary lines in the chain,
and only $q$ of them bear $\tau^*$ operators, hence have "improved" power counting
prefactor $M^{-2r(\si)}  $. The last one has the usual prefactor $ M^{-r(\si) - l(\si )/2}$.  
Multiplying these bounds by the correct factors for the bipeds, namely those in
(\ref{eqimportant}) and (\ref{eqimportant0}), we obtain that the chain is bounded by
\be
\la ^q K^q  M^{-(2q +1)r(\si) - l(\si )/2 }   \int  \prod_{j=1}^{q} dy_j dz_j  
 \prod_{j=0}^q \Big[ e^{-d_{\si} ^{\al}(z_j,y_{j+1})} \Big]  
 \prod_{j=1}^q  \sup_{\bar \si_{R_j,\mathcal{T},j} }  \Big[ e^{-c.d_{R_j,\mathcal{T}_j} ^{\al}(y_j,z_{j} ) }         
M ^{-2r_{R_j,\mathcal{T}_j}} \Big]\ . \label{integall}
\ee
But the integration over all internal points precisely compensates all factors
except the best one. So we can define 
\be i_c = \max  \Big\{ i(\si), \max_j  i_{R_j,\mathcal{T}_j} \Big\}\  ;\   s_{+,c} = \max  \Big\{ s_{+}(\si), \max_j  s_{+,R_j} \Big\}\  ;\  
s_{-,c} = \max  \Big\{ s_{-}(\si), \max_j  s_{-,R_j} \Big\}\  ;\ 
\ee
which can be again considered elements of a triplet $\si_c = \{ i_c,  s_{+,c}, s_{-,c}\}$ ,
with $r_c=  ( i_c + s_{+,c} + s_{-,c})/2$. We obtain 
therefore for the chain $Chain$ with integrated internal vertices  (even after summing over
$q$) again a bound
\be | Chain_r (y,z, \si )  |   \le   K.   \sup_{\si_c \in \Si_r}   M ^{- r_{c} -l(\si)/2 } e^{-c.d_{\si_c} ^{\al}(y,z)}   
\ee
Finally we can check easily by repeating the analysis of (\ref{integall})
that this chain also satisfies the renormalized bounds analogous to 
(\ref{renprop}) and 
(\ref{renproz}):
\be | (\partial_+  + i )(\partial_-  + i )
Chain_r (y,z, \si )  |   \le   K.   \sup_{\bar \si_c \in \Si_r}   M ^{-2r_{c} -l(\si)/2 } e^{-c.d_{\si_c} ^{\al}(y,z)}   
\ee
\be |  (\partial_0   - i\pi T )  Chain_r (y,z, \si )  |   \le   K.   \sup_{\bar \si_c \in \Si_r}   M ^{-2r_{c} } e^{-c.d_{\si_c} ^{\al}(y,z)}   
\ee
Hence any chain of primitively divergent bipeds satisfies to the same
bounds as that of an ordinary line, except that for the logarithmic sum concerning
the momentum conservation of sectors we should use the old sector
$\si$, but for the power counting we should use the prefactor and decay
corresponding to $\si_c$. Plugging this input into an inductive 
argument, we obtain that Theorem \ref{theortwopoint} also holds for 
the sum over all contributions, not necessarily only the primitively divergent ones.

\section{Self-energy Bounds}

In this section we summarize what has been achieved by the previous sections
into  two bounds on the self-energy, one with the first 
non trivial graph $G_2$
included and the other with that graph excluded\footnote{This first non trivial graph is the elementary biped in the chain of Figure 1, since the
tadpole vanishes.}. We apply the analysis 
above to $\Si^{ r} (k)$, the sum of all self-energy contributions 
with lines of index $\le r $ and at least one line of index $r$. In this way, keeping track of the maximal number of $| \log T |$ factors due to quadrupeds in the clustering tree structure, which is $n-1$
at order $n$, it is tedious but straightforward to obtain the following bounds analogs to Theorem \ref{theortwopoint}:

\be | \Si^{r} (k)| \le K  \big(\la | \log T |\big) ^2  M^{-r}  
\ee
\be \left| \frac{\partial}{\partial k_\mu } \Si^{ r} (k) \right| \le  K \big(\la | \log T |\big) ^2   
\ee
\be \left| \frac{\partial^2}{\partial k_\mu \partial k_\nu}  \Si^{r} (k) \right| \le K  \big(\la | \log T |\big) ^2     M^{r}  
\ee
where $K$ is some constant.
The same quantities but with the particular graph $G_2$ taken out give similar 
but slightly better bounds since the series start with contributions of order 3:

\be  | \Si^{r}_{n \ge 3} (k)|  \le  K \la^3 | \log T |^4  M^{-r}  
\ee
\be  \left| \frac{\partial}{\partial k_\mu } \Si^{r}_{n \ge 3} (k) \right|  \le K  \la^3  | \log T |^4     
\ee
\be  \left|  \frac{\partial^2}{\partial k_\mu \partial k_\nu}\Si^{ r}_{n \ge 3} (k)\right|  \le   K  \la^3  | \log T |^4      M^{+r}  
\label{upperb}
\ee

Therefore, one can sum over $r$ the self energy contributions $\Si^ {r} (k)$, $\Si^{r}_{n \ge 3} (k)$
and their first momentum derivatives in the domain $\la |\log T |^2 \le c$ for small $c$, obtaining the bounds
\be | \Si  (k)| \le K  c^2  | \log T |^{-2}    
\ee
\be \left| \frac{\partial}{\partial k_\mu } \Si  (k) \right| \le  K c^2  | \log T | ^{-1}  
\ee
\be | \Si _{n \ge 3} (k)| \le K c^3  | \log T |^{-2}       
\ee
\be  \left| \frac{\partial}{\partial k_\mu } \Si _{n \ge 3} (k) \right| \le  K c^3  | \log T | ^{-1}  
\ee

This proves that the self energy is uniformly $C^1$ in the domain of analyticity
of the theory, namely $\la |\log T |^2 \le c$. However the bounds
for second derivatives grow with $r$, strongly suggesting that
the self-energy is  {\it not} uniformly of class $C^2$
in the domain $|\lambda| \le  c/|\log T|^{2}$, just like the Luttinger liquid
in one dimension.

More precisely we have proved by a tedious analysis (\cite{AMR1}) the following lower bound for the 
amplitude of the single graph $G_2$ 
\be \left| \frac{\partial^2}{\partial k_\mu \partial k_\nu}  \Si_{n=2} (k) \right| =  \left| I_{G_2} (k) \right|
 \ge  K' \la^2 |\log T|^{2}  M^{+r} \ , \label{lower}
\ee
in the special case of $\mu, \nu$ in the $(+,+)$ 
direction and incoming momentum $(k_0 = \pi T, k_+ = 1, k_- =0)$.

This completes the proof that the Hubbard model at half-filling is
{\it not} a Fermi liquid in the sense of \cite{S1}. Indeed for $\la |\log T |^2 \le c$ and $c$ smaller than
$K'/2K$ , the rest of the series, bounded in (\ref{upperb}) by $K  \la^3  | \log T |^4  M^{+r}$, hence by  
$ K c \la^2  | \log T |^2      M^{+r}$, is smaller than half the right hand side of 
(\ref{lower}). When we add it and take $M^{+r} \simeq M^{+r_{max}} = 1/T$,
the modulus of the full quantity  $\frac{\partial^2}{\partial k_\mu \partial k_\nu}  \Si (k)$
therefore diverges at least as $ K' c^2 |\log T|^{-2} /2T $ along the curve $\la |\log T |^2 = c$
as $T\to 0$, which means that Salmhofer's criterion for Ferrmi liquids is violated.

\medskip
\noindent {\bf Acknowledgments}

We thank A. Abdesselam for useful discussions on Menger's theorem.

\end{document}

%% file: arbre.pstex_t
\begin{picture}(0,0)%
\includegraphics{arbre.pstex}%
\end{picture}%
\setlength{\unitlength}{4144sp}%
\begingroup\makeatletter\ifx\SetFigFont\undefined%
\gdef\SetFigFont#1#2#3#4#5{%
  \reset@font\fontsize{#1}{#2pt}%
  \fontfamily{#3}\fontseries{#4}\fontshape{#5}%
  \selectfont}%
\fi\endgroup%
\begin{picture}(6434,2792)(232,-2308)
\put(2215,328){\makebox(0,0)[lb]{\smash{{\SetFigFont{12}{14.4}{\rmdefault}{\mddefault}{\updefault}{\color[rgb]{0,0,0}$\mathfrak{F}_2$}%
}}}}
\put(1253,-1209){\makebox(0,0)[lb]{\smash{{\SetFigFont{12}{14.4}{\rmdefault}{\mddefault}{\updefault}{\color[rgb]{0,0,0}$\mathfrak{F_1}$}%
}}}}
\put(561,136){\makebox(0,0)[lb]{\smash{{\SetFigFont{12}{14.4}{\rmdefault}{\mddefault}{\updefault}{\color[rgb]{0,0,0}$\mathfrak{F}_0$}%
}}}}
\put(5896,-241){\makebox(0,0)[lb]{\smash{{\SetFigFont{12}{14.4}{\familydefault}{\mddefault}{\updefault}{\color[rgb]{0,0,0}$\mathfrak{F}_p$}%
}}}}
\end{picture}%

%% file: arbre2.pstex_t
\begin{picture}(0,0)%
\includegraphics{arbre2.pstex}%
\end{picture}%
\setlength{\unitlength}{3947sp}%
\begingroup\makeatletter\ifx\SetFigFont\undefined%
\gdef\SetFigFont#1#2#3#4#5{%
  \reset@font\fontsize{#1}{#2pt}%
  \fontfamily{#3}\fontseries{#4}\fontshape{#5}%
  \selectfont}%
\fi\endgroup%
\begin{picture}(12519,8241)(204,-8269)
\put(2101,-6136){\makebox(0,0)[lb]{\smash{\SetFigFont{34}{40.8}{\familydefault}{\mddefault}{\updefault}{\color[rgb]{0,0,0}$\mathfrak{F}_1$}%
}}}
\put(11326,-4411){\makebox(0,0)[lb]{\smash{\SetFigFont{34}{40.8}{\familydefault}{\mddefault}{\updefault}{\color[rgb]{0,0,0}$\mathfrak{F}_p$}%
}}}
\put(3526,-3136){\makebox(0,0)[lb]{\smash{\SetFigFont{34}{40.8}{\familydefault}{\mddefault}{\updefault}{\color[rgb]{0,0,0}$\mathfrak{F}_2$}%
}}}
\put(1051,-3436){\makebox(0,0)[lb]{\smash{\SetFigFont{34}{40.8}{\familydefault}{\mddefault}{\updefault}{\color[rgb]{0,0,0}$\mathfrak{F}_0$}%
}}}
\put(3676,-436){\makebox(0,0)[lb]{\smash{\SetFigFont{34}{40.8}{\familydefault}{\mddefault}{\updefault}{\color[rgb]{0,0,0}$\pmb{\ell}_1$}%
}}}
\put(6226,-4861){\makebox(0,0)[lb]{\smash{\SetFigFont{34}{40.8}{\familydefault}{\mddefault}{\updefault}{\color[rgb]{0,0,0}$\mathfrak{F}_{k_1}$}%
}}}
\end{picture}

%% file: arbre3.pstex_t
\begin{picture}(0,0)%
\includegraphics{arbre3.pstex}%
\end{picture}%
\setlength{\unitlength}{3947sp}%
\begingroup\makeatletter\ifx\SetFigFont\undefined%
\gdef\SetFigFont#1#2#3#4#5{%
  \reset@font\fontsize{#1}{#2pt}%
  \fontfamily{#3}\fontseries{#4}\fontshape{#5}%
  \selectfont}%
\fi\endgroup%
\begin{picture}(12519,8466)(204,-8269)
\put(2026,-6211){\makebox(0,0)[lb]{\smash{\SetFigFont{34}{40.8}{\familydefault}{\mddefault}{\updefault}{\color[rgb]{0,0,0}$\mathfrak{F}_1$}%
}}}
\put(826,-3361){\makebox(0,0)[lb]{\smash{\SetFigFont{34}{40.8}{\familydefault}{\mddefault}{\updefault}{\color[rgb]{0,0,0}$\mathfrak{F}_0$}%
}}}
\put(3676,-2911){\makebox(0,0)[lb]{\smash{\SetFigFont{34}{40.8}{\familydefault}{\mddefault}{\updefault}{\color[rgb]{0,0,0}$\mathfrak{F}_2$}%
}}}
\put(11401,-4336){\makebox(0,0)[lb]{\smash{\SetFigFont{34}{40.8}{\familydefault}{\mddefault}{\updefault}{\color[rgb]{0,0,0}$\mathfrak{F}_p$}%
}}}
\put(3226,-211){\makebox(0,0)[lb]{\smash{\SetFigFont{34}{40.8}{\familydefault}{\mddefault}{\updefault}{\color[rgb]{0,0,0}$\pmb{\ell}_1$}%
}}}
\put(6301,-3886){\makebox(0,0)[lb]{\smash{\SetFigFont{34}{40.8}{\familydefault}{\mddefault}{\updefault}{\color[rgb]{0,0,0}$\pmb{\ell}_2$}%
}}}
\end{picture}

%% file: arbre4.pstex_t
\begin{picture}(0,0)%
\includegraphics{arbre4.pstex}%
\end{picture}%
\setlength{\unitlength}{3947sp}%
\begingroup\makeatletter\ifx\SetFigFont\undefined%
\gdef\SetFigFont#1#2#3#4#5{%
  \reset@font\fontsize{#1}{#2pt}%
  \fontfamily{#3}\fontseries{#4}\fontshape{#5}%
  \selectfont}%
\fi\endgroup%
\begin{picture}(12519,8722)(204,-8273)
\put(901,-3286){\makebox(0,0)[lb]{\smash{{\SetFigFont{34}{40.8}{\familydefault}{\mddefault}{\updefault}{\color[rgb]{0,0,0}$\mathfrak{F}_0$}%
}}}}
\put(2026,-6061){\makebox(0,0)[lb]{\smash{{\SetFigFont{34}{40.8}{\familydefault}{\mddefault}{\updefault}{\color[rgb]{0,0,0}$\mathfrak{F}_1$}%
}}}}
\put(11251,-4336){\makebox(0,0)[lb]{\smash{{\SetFigFont{34}{40.8}{\familydefault}{\mddefault}{\updefault}{\color[rgb]{0,0,0}$\mathfrak{F}_p$}%
}}}}
\put(5176,-1786){\makebox(0,0)[lb]{\smash{{\SetFigFont{34}{40.8}{\familydefault}{\mddefault}{\updefault}{\color[rgb]{0,0,0}$\pmb{\ell}_1$}%
}}}}
\put(7051,-586){\makebox(0,0)[lb]{\smash{{\SetFigFont{34}{40.8}{\familydefault}{\mddefault}{\updefault}{\color[rgb]{0,0,0}$\pmb{\ell}_2$}%
}}}}
\end{picture}%

%% file: arbre5.pstex_t
\begin{picture}(0,0)%
\includegraphics{arbre5.pstex}%
\end{picture}%
\setlength{\unitlength}{3947sp}%
\begingroup\makeatletter\ifx\SetFigFont\undefined%
\gdef\SetFigFont#1#2#3#4#5{%
  \reset@font\fontsize{#1}{#2pt}%
  \fontfamily{#3}\fontseries{#4}\fontshape{#5}%
  \selectfont}%
\fi\endgroup%
\begin{picture}(12519,8991)(204,-8269)
\put(4951,-1561){\makebox(0,0)[lb]{\smash{\SetFigFont{34}{40.8}{\familydefault}{\mddefault}{\updefault}{\color[rgb]{0,0,0}$\pmb{\ell}_1$}%
}}}
\put(6601,-3661){\makebox(0,0)[lb]{\smash{\SetFigFont{34}{40.8}{\familydefault}{\mddefault}{\updefault}{\color[rgb]{0,0,0}$\pmb{\ell}_2$}%
}}}
\put(5251,314){\makebox(0,0)[lb]{\smash{\SetFigFont{34}{40.8}{\familydefault}{\mddefault}{\updefault}{\color[rgb]{0,0,0}$\pmb{\ell}_3$}%
}}}
\end{picture}

%% file: 3types.pstex_t
\begin{picture}(0,0)%
\includegraphics{3types.pstex}%
\end{picture}%
\setlength{\unitlength}{3947sp}%
\begingroup\makeatletter\ifx\SetFigFont\undefined%
\gdef\SetFigFont#1#2#3#4#5{%
  \reset@font\fontsize{#1}{#2pt}%
  \fontfamily{#3}\fontseries{#4}\fontshape{#5}%
  \selectfont}%
\fi\endgroup%
\begin{picture}(12494,7113)(204,-7744)
\end{picture}

%% file: arbre6.pstex_t
\begin{picture}(0,0)%
\includegraphics{arbre6.pstex}%
\end{picture}%
\setlength{\unitlength}{3947sp}%
\begingroup\makeatletter\ifx\SetFigFont\undefined%
\gdef\SetFigFont#1#2#3#4#5{%
  \reset@font\fontsize{#1}{#2pt}%
  \fontfamily{#3}\fontseries{#4}\fontshape{#5}%
  \selectfont}%
\fi\endgroup%
\begin{picture}(12519,8332)(204,-8273)
\put(3526,-361){\makebox(0,0)[lb]{\smash{{\SetFigFont{34}{40.8}{\familydefault}{\mddefault}{\updefault}{\color[rgb]{0,0,0}$\pmb{\ell}_1$}%
}}}}
\end{picture}%

%% file: arbre7.pstex_t
\begin{picture}(0,0)%
\includegraphics{arbre7.pstex}%
\end{picture}%
\setlength{\unitlength}{3947sp}%
\begingroup\makeatletter\ifx\SetFigFont\undefined%
\gdef\SetFigFont#1#2#3#4#5{%
  \reset@font\fontsize{#1}{#2pt}%
  \fontfamily{#3}\fontseries{#4}\fontshape{#5}%
  \selectfont}%
\fi\endgroup%
\begin{picture}(12519,8595)(204,-8536)
\put(3226,-8536){\makebox(0,0)[lb]{\smash{{\SetFigFont{20}{24.0}{\familydefault}{\mddefault}{\updefault}{\color[rgb]{0,0,0}cuts}%
}}}}
\put(3676,-361){\makebox(0,0)[lb]{\smash{{\SetFigFont{34}{40.8}{\familydefault}{\mddefault}{\updefault}{\color[rgb]{0,0,0}$\pmb{\ell}_1$}%
}}}}
\put(5926,-2761){\makebox(0,0)[lb]{\smash{{\SetFigFont{34}{40.8}{\familydefault}{\mddefault}{\updefault}{\color[rgb]{0,0,0}$\pmb{\ell}_2$}%
}}}}
\end{picture}%

%% file: numbering.pstex_t
\begin{picture}(0,0)%
\includegraphics{numbering.pstex}%
\end{picture}%
\setlength{\unitlength}{3947sp}%
\begingroup\makeatletter\ifx\SetFigFont\undefined%
\gdef\SetFigFont#1#2#3#4#5{%
  \reset@font\fontsize{#1}{#2pt}%
  \fontfamily{#3}\fontseries{#4}\fontshape{#5}%
  \selectfont}%
\fi\endgroup%
\begin{picture}(13299,3970)(-11,-7157)
\put(3526,-3586){\makebox(0,0)[lb]{\smash{\SetFigFont{29}{34.8}{\rmdefault}{\mddefault}{\updefault}{\color[rgb]{0,0,0}$\mathfrak{G}_{1,a}$}%
}}}
\put(9151,-3511){\makebox(0,0)[lb]{\smash{\SetFigFont{29}{34.8}{\rmdefault}{\mddefault}{\updefault}{\color[rgb]{0,0,0}$\mathfrak{G}_{3,b}$}%
}}}
\put(12676,-4861){\makebox(0,0)[lb]{\smash{\SetFigFont{29}{34.8}{\rmdefault}{\mddefault}{\updefault}{\color[rgb]{0,0,0}$\mathfrak{G}_{5}$}%
}}}
\put(5701,-6361){\makebox(0,0)[lb]{\smash{\SetFigFont{29}{34.8}{\rmdefault}{\mddefault}{\updefault}{\color[rgb]{0,0,0}$\mathfrak{G}_{2,a'}$}%
}}}
\put(8326,-6961){\makebox(0,0)[lb]{\smash{\SetFigFont{29}{34.8}{\rmdefault}{\mddefault}{\updefault}{\color[rgb]{0,0,0}$\mathfrak{G}_{3,b'}$}%
}}}
\put(9601,-7036){\makebox(0,0)[lb]{\smash{\SetFigFont{29}{34.8}{\rmdefault}{\mddefault}{\updefault}{\color[rgb]{0,0,0}$\mathfrak{G}_{3,c'}$}%
}}}
\put(7201,-6361){\makebox(0,0)[lb]{\smash{\SetFigFont{29}{34.8}{\rmdefault}{\mddefault}{\updefault}{\color[rgb]{0,0,0}$\mathfrak{G}_{3,a'}$}%
}}}
\put(3376,-7036){\makebox(0,0)[lb]{\smash{\SetFigFont{29}{34.8}{\rmdefault}{\mddefault}{\updefault}{\color[rgb]{0,0,0}$\mathfrak{G}_{1,a'}$}%
}}}
\put(226,-4711){\makebox(0,0)[lb]{\smash{\SetFigFont{29}{34.8}{\rmdefault}{\mddefault}{\updefault}{\color[rgb]{0,0,0}$\mathfrak{G}_{0}$}%
}}}
\put(751,-4186){\makebox(0,0)[lb]{\smash{\SetFigFont{29}{34.8}{\rmdefault}{\mddefault}{\updefault}{\color[rgb]{0,0,0}$\mathfrak{G}_{0,a}$}%
}}}
\put(6226,-4636){\makebox(0,0)[lb]{\smash{\SetFigFont{29}{34.8}{\rmdefault}{\mddefault}{\updefault}{\color[rgb]{0,0,0}$\mathfrak{G}_{3}$}%
}}}
\put(2626,-5311){\makebox(0,0)[lb]{\smash{\SetFigFont{29}{34.8}{\rmdefault}{\mddefault}{\updefault}{\color[rgb]{0,0,0}$\mathfrak{G}_{1}$}%
}}}
\put(5026,-5236){\makebox(0,0)[lb]{\smash{\SetFigFont{29}{34.8}{\rmdefault}{\mddefault}{\updefault}{\color[rgb]{0,0,0}$\mathfrak{G}_{2}$}%
}}}
\put(7276,-3811){\makebox(0,0)[lb]{\smash{\SetFigFont{29}{34.8}{\rmdefault}{\mddefault}{\updefault}{\color[rgb]{0,0,0}$\mathfrak{G}_{3,a}$}%
}}}
\put(11326,-5386){\makebox(0,0)[lb]{\smash{\SetFigFont{29}{34.8}{\rmdefault}{\mddefault}{\updefault}{\color[rgb]{0,0,0}$\mathfrak{G}_{4}$}%
}}}
\end{picture}

%% file: twopipoisson.pstex_t
\begin{picture}(0,0)%
\includegraphics{twopipoisson.pstex}%
\end{picture}%
\setlength{\unitlength}{3947sp}%
\begingroup\makeatletter\ifx\SetFigFont\undefined%
\gdef\SetFigFont#1#2#3#4#5{%
  \reset@font\fontsize{#1}{#2pt}%
  \fontfamily{#3}\fontseries{#4}\fontshape{#5}%
  \selectfont}%
\fi\endgroup%
\begin{picture}(15770,11564)(-2622,-10848)
\put(976,-2836){\makebox(0,0)[lb]{\smash{\SetFigFont{29}{34.8}{\rmdefault}{\mddefault}{\updefault}{\color[rgb]{0,0,0}$\pmb{\ell}_2$}%
}}}
\put(3751,-4411){\makebox(0,0)[lb]{\smash{\SetFigFont{29}{34.8}{\rmdefault}{\mddefault}{\updefault}{\color[rgb]{0,0,0}$\pmb{\ell}_3$}%
}}}
\put(1126,-7786){\makebox(0,0)[lb]{\smash{\SetFigFont{29}{34.8}{\rmdefault}{\mddefault}{\updefault}{\color[rgb]{0,0,0}$\pmb{\ell}_1$}%
}}}
\put(11626,-2311){\makebox(0,0)[lb]{\smash{\SetFigFont{29}{34.8}{\rmdefault}{\mddefault}{\updefault}{\color[rgb]{0,0,0}$\pmb{\ell}_6$}%
}}}
\put(6601,-2161){\makebox(0,0)[lb]{\smash{\SetFigFont{29}{34.8}{\rmdefault}{\mddefault}{\updefault}{\color[rgb]{0,0,0}$\pmb{\ell}_5$}%
}}}
\put(6976,-7711){\makebox(0,0)[lb]{\smash{\SetFigFont{29}{34.8}{\rmdefault}{\mddefault}{\updefault}{\color[rgb]{0,0,0}$\pmb{\ell}_4$}%
}}}
\end{picture}